\documentclass[amsmath,twocolumn,prx,aps]{revtex4-1} 
\usepackage{amsthm,amsfonts,graphicx,verbatim, color}
\usepackage{bm}
\usepackage[utf8]{inputenc}
\let\oldmarginpar\marginpar
\renewcommand\marginpar[1]{\-\oldmarginpar[\raggedleft\footnotesize #1]%
{\raggedright\footnotesize #1}}

\newcommand{\be}{\begin{equation}}
\newcommand{\ee}{\end{equation}}
\newcommand{\bea}{\begin{eqnarray}}
\newcommand{\eea}{\end{eqnarray}}

\newcommand{\Tr}{{\rm Tr}\,}
\renewcommand{\epsilon}{\varepsilon}

\def\beq{\begin{equation}}
\def\eeq{\end{equation}}
\def\bea{\begin{eqnarray}}
\def\eea{\end{eqnarray}}

\begin{document}

\title{Many body localization and thermalization: insights from the entanglement spectrum}

\author{Scott D. Geraedts}
\affiliation{Department of Electrical Engineering, Princeton University, Princeton NJ 08544, USA}
\author{Rahul Nandkishore}
\affiliation{Department of Physics, University of Colorado, Boulder, Colorado 80309, USA}
\affiliation{Center for Theory of Quantum Matter, University of Colorado, Boulder, Colorado 80309, USA}
\author{Nicolas Regnault}
\affiliation{Department of Physics, Princeton University, Princeton NJ 08544, USA}
\affiliation{Laboratoire Pierre Aigrain, Ecole Normale Sup\'erieure-PSL Research
University, CNRS, Universit\'e Pierre et Marie Curie-Sorbonne Universit\'es,
Universit\'e Paris Diderot-Sorbonne Paris Cit\'e, 24 rue Lhomond, 75231
Paris Cedex 05, France}

\begin{abstract}
We study the entanglement spectrum in the many body localizing and thermalizing phases of one and two dimensional Hamiltonian systems, and periodically driven `Floquet' systems. We focus on the level statistics of the entanglement spectrum as obtained through numerical diagonalization, finding structure beyond that revealed by more limited measures such as entanglement entropy. In the thermalizing phase the entanglement spectrum obeys level statistics governed by an appropriate random matrix ensemble. For Hamiltonian systems this can be viewed as evidence in favor of a strong version of the eigenstate thermalization hypothesis (ETH). Similar results are also obtained for Floquet systems, where they constitute a result `beyond ETH', and show that the corrections to ETH governing the Floquet entanglement spectrum have statistical properties governed by a random matrix ensemble. The particular random matrix ensemble governing the Floquet entanglement spectrum depends on the symmetries of the Floquet drive, and therefore can depend on the choice of origin of time. In the many body localized phase the entanglement spectrum is also found to show level repulsion, following a semi-Poisson distribution (in contrast to the energy spectrum, which follows a Poisson distribution). This semi-Poisson distribution is found to come mainly from states at high entanglement energies. The observed level repulsion only occurs for interacting localized phases.
We also demonstrate that equivalent results can be obtained by calculating with a single typical eigenstate, or by averaging over a microcanonical energy window - a surprising result in the localized phase. This discovery of new structure in the pattern of entanglement of localized and thermalizing phases may open up new lines of attack on many body localization, thermalization, and the localization transition. 
\end{abstract}
\maketitle

\section{Introduction}

Recent advances in the synthesis and control of well isolated quantum systems, and the increasing importance of such systems as building blocks for new quantum technologies, have triggered an explosion of interest in the statistical mechanics of closed many body quantum systems. Particular interest has centered on the question of whether (and how) isolated many body quantum systems go to thermal equilibrium, and on the validity of the ergodic hypothesis. It appears that two generic behaviors are possible for isolated quantum systems, depending on system details. Either systems can {\it thermalize}, with local observables approaching equilibrium expectation values and the ergodic hypothesis being satisfied, or they can {\it localize}, displaying non-ergodic behavior and failing to reach thermal equilibrium even at infinite times \cite{Nandkishore-2015}. Thermalizing systems have been largely studied through the lens of the Eigenstate Thermalization Hypothesis (ETH) \cite{Deutsch, Srednicki, Rigol}, which explains how equilibrium statistical mechanics can continue to hold in an isolated quantum system. The ETH is violated by {\it many body localized} systems, which exhibit a wealth of non-equilibrium phenomena, including an emergent integrability \cite{Serbynlbits, lbits}, exotic quantum phases with no analog in equilibrium \cite{LPQO, Pekkeretal, VoskAltman, Bauer, Bahri, Chandran}, a non-local response to local perturbations \cite{nonlocal}, and unusual scaling of response functions \cite{response}. An understanding is rapidly emerging of when systems {\it can} \cite{agkl, Gornyi, BAA, oganesyan, Pal, Imbrie} or can not \cite{burin, QHMBL, yaodipoles, 2dcontinuum} be localized, but our understanding of the localized and delocalized phases themselves, as well as the transition between them, remains a work in progress. 

Ideas from {\it quantum information} have played a central role in the developing understanding of localization and thermalization. Particularly fruitful has been the notion of {\it quantum entanglement}, as characterized by {\it entanglement entropy}. Many body eigenstates of thermalizing systems exhibit an entanglement entropy that scales with the {\it volume} of the subregion being considered, while many body eigenstates of many body localized systems display a boundary law scaling (with possible logarithmic corrections). The study of entanglement entropy and its dynamics has played a crucial role in the eludication of the properties of the localized and thermal phases. Indeed studies of entanglement entropy \cite{Bardarson} provided the first clues as to the emergent integrability of the localized phase, and have also been used to constrain the properties of the localization transition \cite{Grover}. However, entanglement entropy captures only a small part of the full entanglement structure of a system. Much greater information is contained in the {\it entanglement spectrum} \cite{LiHaldane}, from which the entanglement entropy may be extracted, and much besides. However, studies of the localizing and thermalizing phase have yet to exploit the wealth of information contained in the entanglement spectrum. 

In this work, we track the evolution of the entanglement spectrum across the many body localization-delocalization transition, and use it to extract new insights into the many body localizing and thermalizing phases. We unlock the information contained in the entanglement spectrum by applying ideas from {\it random matrix theory}, which provides a powerful mechanism for understanding thermalization, complementary to the ETH. In particular, we apply diagnostics from random matrix theory to the level statistics of entanglement spectra. Previous investigations of level statistics \cite{Pal} focused on the energy spectrum. While a recent work \cite{Chamon} did examine the entanglement spectrum, it focused mostly on the density of states distribution, and on the thermal phase of Hamiltonian systems. We focus instead on the level statistics, and examine both the thermal and the many body localized phase, in both Hamiltonian and Floquet systems. The structure revealed, including remnants of criticality in the entanglement structure deep in the localized phase, and features beyond ETH in the entanglement structure of thermal states, have never before been seen, as far as we are aware. 

This work is structured as follows. We begin by investigating the entanglement structure across the many body localization transition in {\it Hamiltonian} systems (i.e. systems with a conserved energy). Performing numerical exact diagonalization on a spin-$1/2$ Heisenberg chain in external field (the canonical model for studies of the transition), we find that in the thermal phase, the entanglement spectrum exhibits level statistics that match the predictions of random matrix theory, and are governed by the same random matrix ensemble as the energy spectrum. This may be viewed as evidence in support of a particularly strong version of the ETH \cite{garrisongrover}, insofar as the usual formulation of the ETH predicts that the entanglement spectrum of a small subregion should obey the same level statistics as the energy spectrum \cite{representativestates}, but we observe this to be true even for subregions that are half the size of the system. The level statistics of the entanglement spectrum are found to be independent of the choice of eigenstate. 

Next, we consider the entanglement spectrum in the localized phase. We discover that in the localized phase, the entanglement spectrum differs substantially from the energy spectrum, and fits to a {\it semi-Poisson} form that is generally considered indicative of {\it criticality} \cite{Serbyn}. This `critical' structure in the entanglement spectrum (and in particular the existence of entanglement level repulsion even deep in the localized phase where the energy spectrum shows no level repulsion) constitutes a new and unexpected property of the localized phase. The statistical properties of the entanglement spectrum appear to be a universal feature that varies only slowly from one eigenstate to the next, such that microcanonical averages of level statistics parameters give equivalent results to calculations with single eigenstates, even though there is no ETH given reason for this to be true in the localized phase. 

A detailed analysis of the entanglement spectrum reveals that these signatures of criticality in the entanglement structure are restricted to states at {\it high} entanglement energies, with the region of the entanglement spectrum at low entanglement energy showing no signs of entanglement level repulsion. Since entanglement entropy and Renyi entropies are dominated by low entanglement energies, the residual criticality in the entanglement structure is likely invisible to these entropy measures, and could only have been revealed by an investigation of the full entanglement spectrum, explaining why this structure has never before been observed. We develop a picture whereby at the critical point, the entire entanglement spectrum is critical, but as the system moves into the localized phase, the low entanglement energies become non-critical, erasing all signatures of criticality in standard diagnostics such as entanglement entropy. However, the high entanglement energy region of the spectrum `sticks' at criticality, such that the entanglement structure of eigenstates continues to contain a memory of criticality, albeit one inaccessible to entropy measures. 

We conjecture an explanation for the residual criticality of the entanglement structure deep in the localized phase as being due to {\it many body resonances}. This explanation predicts that the observed residual criticality is a consequence of the {\it interacting} nature of the many body localized phase, and would be absent in the non-interacting Anderson insulator. A numerical investigation of $XXZ$ spin chains provides strong numerical support for this scenario, uncovering a new and hitherto unsuspected distinction between the entanglement structure of single particle and many body localized eigenstates. 

We note that semi-Poisson statistics are a known diagnostic of {\it pseudo-integrability} \cite{Berry, Bogomolny}, and our results this suggest that while the entanglement Hamiltonian of a non-interacting Anderson insulator will be integrable, the entanglement Hamiltonian of a many body localized system will be only pseudo-integrable (i.e. not chaotic but also not integrable). This observation may open new lines of attack on the many body localized phase. 

The investigation in the first part of the paper worked with the `canonical' model of a one dimensional spin chain. Next, we study the evolution of the entanglement spectrum across the many body localization-delocalization transition in a {\it two dimensional} model, of a transverse field Ising spin chain. Again, we observe that the level splittings in the thermal phase follow predictions of random matrix theory. Since this two dimensional model can thermalize even in the absence of disorder (whereas the canonical one dimensional models are integrable in the absence of disorder), we can confirm that the observed level statistics of the entanglement spectrum are purely a property of {\it thermalizing} states, and do not depend on the presence or absence of disorder. Upon turning on disorder (in the form of random magnetic fields) and driving the system across the localization transition, we find that the entanglement spectrum evolves in precisely the same way as in one dimensional systems, displaying semi-Poisson level statistics and residual criticality deep into the localized phase. This provides strong numerical evidence that the entanglement structure of higher dimensional many body localized phases is analogous to one dimensional localized phases. We believe this is the first systematic study of the entanglement structure across the many body localization transition in $d>1$. 

In the final section of the paper, we track the evolution of the entanglement spectrum across the many body localization transition in periodically driven {\it`Floquet'} systems. Floquet systems, due to the lack of energy conservation constraints, provide a particularly interesting playground for investigating questions of localization and thermalization, as well as providing a potential new universality class for the transition. We find that the entanglement spectrum in the Floquet localized phase universally obeys semi-Poisson statistics and displays residual criticality, much as in Hamiltonian systems. However, we find more unexpected features in the entanglement spectrum of the Floquet thermalizing phase. 

The level statistics of the spectrum of Floquet eigenphases (the Floquet analog of the energy spectrum) have previously been studied, e.g. in \cite{dAlessio, Regnault, Bukov}, where it was found that in the thermalizing phase Floquet eigenphase spectra obey predictions of random matrix theory, albeit potentially governed by a different ensemble to the instantaneous Hamiltonian \cite{Regnault}. However, unlike in Hamiltonian systems, the {\it entanglement} spectrum is unrelated to the eigenphase spectrum by ETH. This is because ETH relates the entanglement Hamiltonian to $\beta H$, where $H$ is the actual Hamiltonian, but Floquet systems thermalize to infinite temperatures $\beta = 0$, such that ETH is totally non-predictive where the entanglement spectrum is concerned. Indeed, the entanglement spectrum is dominated by {\it corrections to ETH}, and thus provides a powerful probe of {\it thermalization beyond ETH}.

We find that the entanglement spectrum in the Floquet thermal phase continues to exhibit level statistics characteristic of random matrix theory, indicating that random matrix theory is a more powerful tool for the study of thermalization than ETH alone. However, the random matrix ensemble governing the entanglement spectrum of a Floquet system may not be the same as the random matrix ensemble governing the spectrum of Floquet eigenphases. Moreover, the entanglement spectrum of Floquet systems is sensitive to symmetries of the entire Floquet drive, and the governing random matrix ensemble can depend on the choice of the origin of time. This indicates that the entanglement structure of the Floquet `Bloch' states {\it changes} over the period of the drive, such that the entanglement spectrum depends on when in the drive period we choose to probe the system. We discuss how one may deduce the random matrix ensemble that will govern the entanglement spectrum of a Floquet thermalizing system. 

In conclusion, our numerical investigation reveals new and hitherto unsuspected {\it entanglement} structure in both the localized and thermalizing phase. This structure is invisible to entropy measures, and is revealed only by applying ideas from random matrix theory to the level statistics of the full entanglement spectrum. Key (and universal) properties include a residual criticality in the entanglement structure deep in the localized phase, which is absent in the Anderson insulator, the lack of dependence of the results on dimensionality, the equivalence of microcanonical averages of entanglement level statistics parameters to single eigenstate calculations even in the localized phase, and the continuing applicability of random matrix theory to the entanglement spectrum of Floquet thermalizing systems (albeit in a `gauge dependent manner' that is sensitive to the origin of time), even when such applicability is not guaranteed by ETH. 

\section{Energy spectrum statistics in a one dimensional spin chain}

\subsection{The model}

Before we present our results on entanglement spectrum statistics, we first review the statistics of the energy levels. For more detail one can consult an earlier study \cite{Regnault} of the localization-delocalization transition, which contains similar results.
We begin our investigation by considering the canonical model for numerical studies of many body localization and thermalization: a one dimensional system of $N$ spins-$1/2$. We choose to work with the Heisenberg Hamiltonian with a random anisotropic field and periodic boundary conditions 
\begin{eqnarray}
H\left(\{h_{\alpha}\}\right)&=& \sum_{\alpha=x,y,z}\left[J \sum_{i=1}^{N} \left[ S^{\alpha}_i S^{\alpha}_{i + 1}\right]+  h_{\alpha} \sum_{i=1}^{N} c_{\alpha,i} S^{\alpha}_i \right] \label{ModelHamiltonian}
\end{eqnarray}
where $S^{\alpha}_i = \frac12 \sigma^{\alpha}_i$ and $\sigma^{\alpha}$ is a Pauli matrix. The coefficients $c_{\alpha,i}$ are uncorrelated and chosen according to a uniform distribution within the interval $[-1,1]$. The amplitude of the random  field is set through the $h_{\alpha}$. In the following, we set $J=1$. Unlike the canonical model widely used in the literature \cite{Pal}, we do not assume that the random field is necessarily uniaxial. Relaxing this assumption allows us to access a wider range of regimes. If we take $h_z = h$, this corresponds to the canonical model with uniaxial random field \cite{Pal}, which displays a many body localized phase (with Poisson energy level statistics) for large $h \gtrsim 3.5$, and a thermalizing phase with Gaussian orthogonal ensemble (GOE) level statistics in a sector with fixed total $S^z$ for small $h$\cite{Pal}. Note that the level statistics are GOE even though the time reversal symmetry is broken by the  field because of the presence of a disguised antiunitary symmetry, made up of time reversal and a rotation by $\pi$ of all spins about the $x$ axis, which leaves the Hamiltonian unchanged. 

In this work we often find it convenient to work with a system without fixed total $S^z$. 
We therefore consider a system where two components of the field are non-zero (e.g. $h_x \neq 0$, $h_z \neq 0$, $h_y = 0$). Here too the level statistics are described by GOE, for small fields when the system thermalizes. The relevant antiunitary symmetry is now time reversal plus a $\pi$ rotation about the $y$ axis (i.e. $S^x \rightarrow S^x$, $S^y \rightarrow -S^y$, $S^z \rightarrow S^z$), which leaves the Hamiltonian unchanged\cite{Avishai,Modak-2014}. Once all three fields are non-zero however there is no longer any such antiunitary symmetry, and the level statistics in the thermalizing phase are described by the Gaussian unitary ensemble (GUE), and the many body localization transition is associated with a change in the level statistics from Poisson to GUE, instead of Poisson to GOE as in the canonical model.

In a previous work \cite{Regnault} we have found that for the model with $h_x=h_z=h$, $h_y=0$, the transition from thermal to localized energy level statistics occurs around $h\simeq 3$ (i.e. $h_c \sqrt{2} \approx 4.2$), while when $h_x=h_y=h_z = h$ the transition is at $h\simeq 2.5$ (i.e. $h_c \sqrt{3} \approx 4.3$). Thus a localized and thermalizing phase exist in both the orthogonal class without $S^z$ conservation, and in the unitary symmetry class, and there does not appear to be any significant difference in the {\it total} disorder strength for the phase transition. 

\subsection{Energy level statistics across the many body localization transition}

\begin{figure}[htb]
\includegraphics[width=\linewidth]{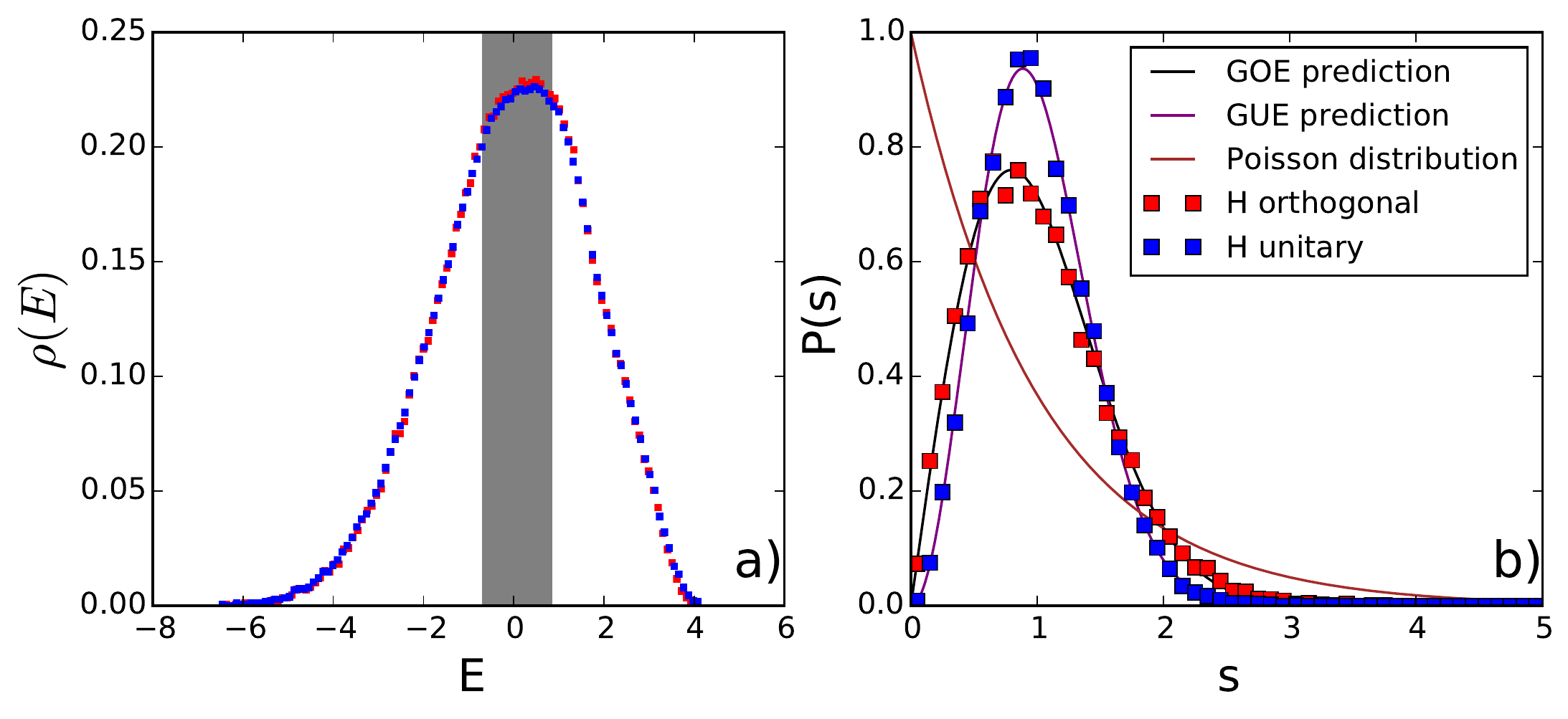}
\caption{ Results from numerical exact diagonalization on the one dimensional spin chain (Eq.\ref{ModelHamiltonian}) showing density of states (a) and level statistics (b) for a system of $N=14$ spins for $h_x=h_z=0.25$ and $h_y=0$ corresponding to GOE (blue dots) or $h_x=h_z=h_y=0.25$ corresponding to GUE (red dots). Even without any average over disorder, we clearly see the good agreement between energy level statistics in the thermal phase and the predictions of random matrix theory for the appropriate ensemble (solid lines). The level statistics clearly differ from the Poisson distribution (brown line). We also show in the left panel the region of the spectrum (shaded area) that we will consider for the entanglement spectrum analysis (we average over the states in the middle third of the spectrum).
}\label{EnergyDOSLevelStatGOEGUE}
\end{figure}

In Figs.~\ref{EnergyDOSLevelStatGOEGUE}a and~\ref{EnergyDOSLevelStatGOEGUE}b we give the density of states and the level statistics (distribution of spacing between levels) for two systems, one in the orthogonal symmetry class without $S^z$ conservation (setting $h_x=h_z=h$ and $h_y=0$) and one in the unitary symmetry class (using $h_x=h_y=h_z=h$). We choose a value of $h$ such that the system is in the thermalized phase. 
A similar study was performed in Ref.~\cite{Regnault}, but here we study larger sizes ($N=14$ for Fig.\ref{EnergyDOSLevelStatGOEGUE}). 
We also present data for a \emph{single} realization of the disorder. Thus we do not perform any average over disorder. 
 Despite looking at a single system with a relatively small number of spins, we clearly observe the level statistics follow predictions of random matrix theory for the appropriate ensemble. The data incorporates the full energy spectrum (not just the middle of the band) using the usual unfolding procedure \cite{Avishai}.

\begin{figure}[htb]
\includegraphics[width=\linewidth]{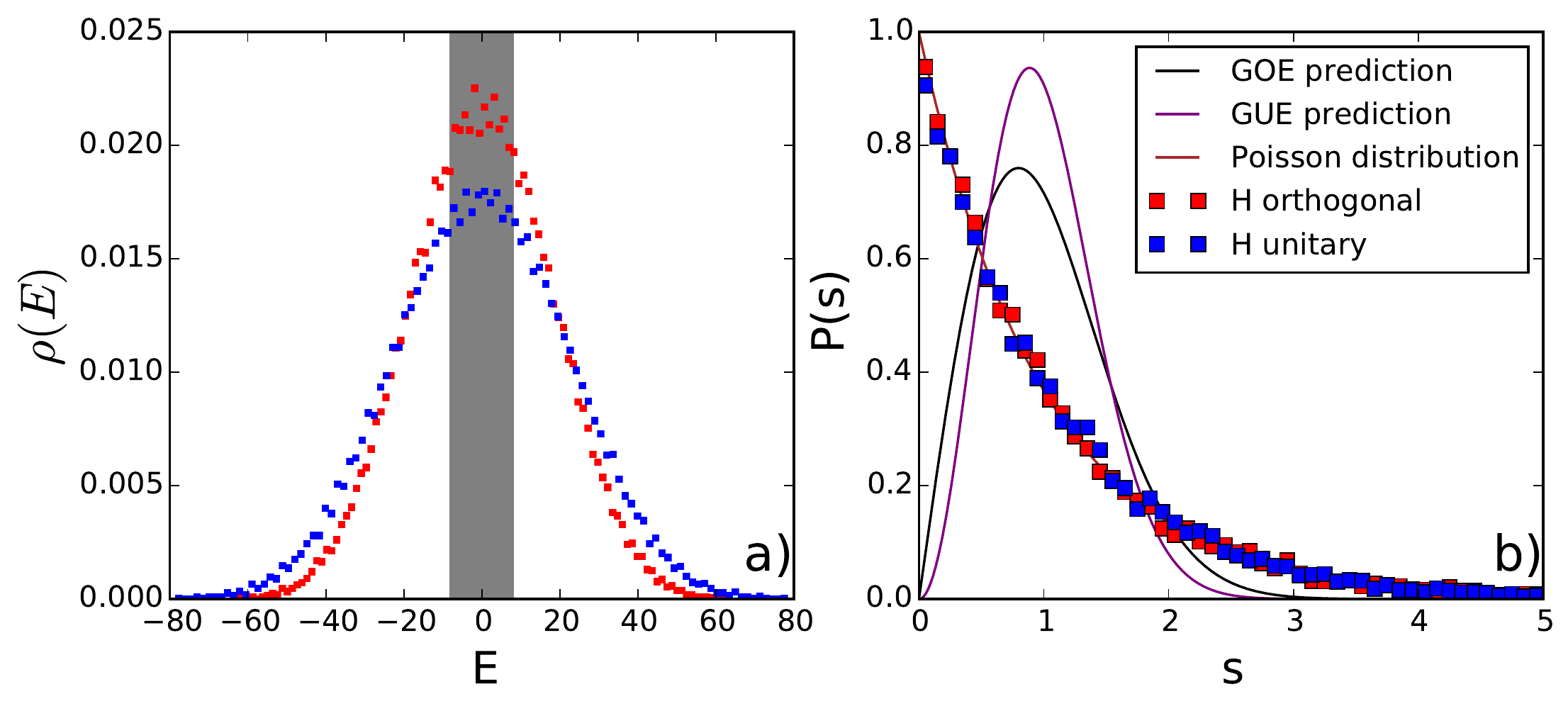}
\caption{Density of states (a) and level statistics (b) for a system of $N=14$ spins for $h_x=h_z=12$ and $h_y=0$ corresponding to the orthogonal class (black dots) or $h_x=h_z=h_y=12$ corresponding to the unitary class (red dots). Even without any average over disorder, we clearly see the good agreement with the Poisson distribution (blue line). We also show in the left panel the region of the spectrum (shaded area) that we will consider for the entanglement spectrum analysis 
(we average over the middle third of states).
}\label{EnergyDOSLevelStatPoisson}
\end{figure}

The energy spectrum of a typical sample deep in the localized phase is summarized in Figs.~\ref{EnergyDOSLevelStatPoisson}a and~\ref{EnergyDOSLevelStatPoisson}b, which show the density of states and the level statistics when $h_x=h_z=12$ and either $h_y=0$ or  $h_y=12$. We clearly observe that the level spacing follows the Poisson distribution, irrespective of whether the Hamiltonian is in the orthogonal or unitary symmetry class.

\section{Entanglement spectrum in the thermalizing phase}\label{Sec:ETHES}

We now come to the main focus of this work: the study of the entanglement spectrum. We begin by studying the entanglement spectrum in the thermalizing phase. We discover that the entanglement spectrum follows the predictions of random matrix theory, and is in the same symmetry class as the energy spectrum. This can be viewed as evidence for a particularly strong version of the ETH, as we shall discuss. 

The entanglement spectrum of an eigenstate can be defined as follows\cite{LiHaldane}: we first consider a system prepared in an eigenstate $|\psi\rangle$. We then cut the system into two subregions $A$ and $B$. Unless otherwise specified, we always consider the `standard entanglement cut' that partitions the system (of $N$ spins) into two equal subregions containing $N/2$ spins each i.e. we cut the system in half. We then construct the reduced density matrix $\rho_A(\psi) = \Tr_B |\psi\rangle \langle \psi|$. Taking the log of the reduced density matrix then defines the {\it entanglement Hamiltonian} $H_{\rm ent}(\psi, A)$ for that eigenstate and cut i.e. 
\begin{equation}
H_{\rm ent}(\psi, A) = - \ln \rho_A(\psi) = - \ln (\Tr_B |\psi\rangle \langle \psi|) \label{entanglementHamiltonian}
\end{equation}
The entanglement spectrum is the spectrum of this entanglement Hamiltonian that we denote ${\xi_i}$. 

While conceptually clean, the above definition is not the most convenient for numerical work. Indeed, it is instead more convenient to consider the Schmidt decomposition of an eigenstate $|\psi\rangle$ into two subregions $A$ and $B$ i.e. 
\begin{equation}
|\psi \rangle = \sum_{i} e^{-\xi_i / 2} |A:i\rangle \otimes |B:i\rangle
\end{equation}
where the $|A:i\rangle$ states (resp. $|B:i\rangle$) form an orthonormal basis restricted to the region $A$ (resp. $B$). Note that $|A:i\rangle$ are also the eigenstates of the entanglement Hamiltonian $H_{\rm ent}(\psi, A)$. The Schmidt decomposition is obtained through singular value decomposition and it provides better numerical accuracy for the $\xi_i$'s, an important advantage when analyzing the high entanglement energy features of the entanglement spectrum.

When studying the entanglement spectrum, we must decide first how to calculate entanglement statistics, since {\it every} many body eigenstate has its own entanglement spectrum, and the number of many body eigenstates is exponentially large in system size even for a single disorder realization. One possible approach is to study the entanglement spectrum for a single typical eigenstate. A drawback of this approach is that the number of entanglement energies per entanglement spectrum is at most $2^{N/2}$. Therefore we will need to study fairly large sizes to extract useful data, and even then the small number of entanglement energies will mean that we incur a large statistical error.

We can reduce these problems by averaging many entanglement spectra corresponding to different eigenstates of the same Hamiltonian. This will allows us to get better statistics even though we are limited to small systems. 
In the ETH phase we expect that every eigenstate should have the properties of the Hamiltonian \cite{representativestates}, and therefore this averaging should not distort our data. 
 Sorting the energies from the smallest to the largest $\{E_i, E_i \le  E_{i+1}\}$, we consider all states between two indices $i_{\rm min}$ and $i_{\rm max}$. These latest are chosen such that we cover a large fraction of spectrum bulk while corresponding to an almost constant density of states. 
Typically we average over all of the states in the middle third of the spectrum following the path trodden by \cite{Pal}. In the Appendix we further justify this choice. 
In Fig.~\ref{EnergyDOSLevelStatGOEGUE}a we shade the region that we have considered for our calculations at $N=14$. We will verify {\it post facto} that the averaged results agree with the results obtained from a single many body eigenstate. 

\begin{figure}
\includegraphics[width=\linewidth]{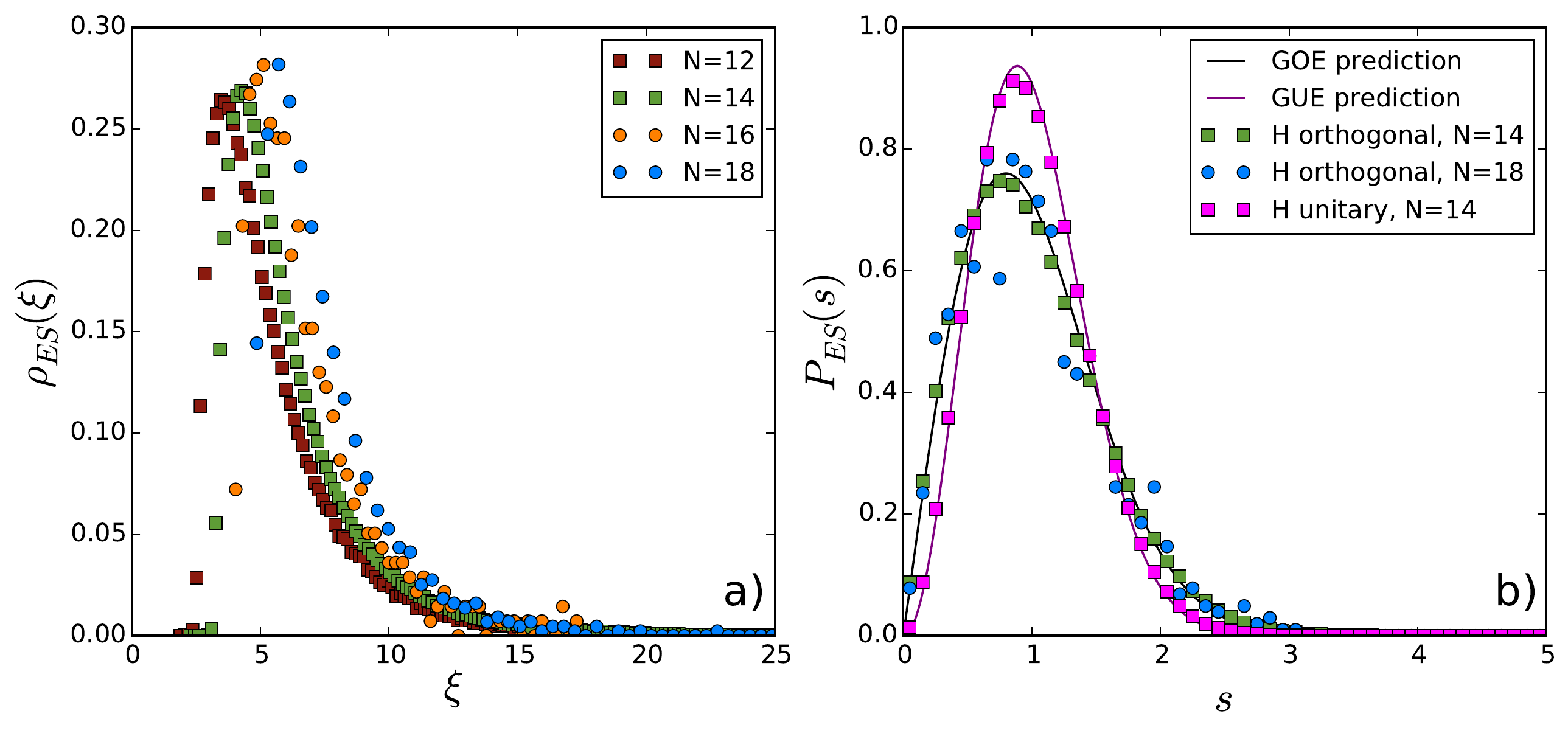}
\caption{(a) Density of states for the entanglement spectrum in the thermalizing phase of (Eq.\ref{ModelHamiltonian}) (with $h_x=h_z=0.25$, $h_y=0$) and for a variety of different system sizes. We work with the standard entanglement cut which cuts the system into two equal halves. Note that the density of states has qualitatively the same behavior whether it is extracted from a single entanglement spectrum ($N=16,18$) or an average of many entanglement spectra ($N=12,14$). We have obtained qualitatively similar data for the unitary Hamiltonian.
(b) Level spacing distributions for the entanglement spectrum in the thermal phase of Eq.\ref{ModelHamiltonian} ($h_x=h_z=0.25$) for chains obeying GOE ($h_y=0$) and GUE($h_y=0.25$). We work with the standard entanglement cut that cuts the system into two equal halves. 
The solid lines show the corresponding predictions for the energy spacings distributions, and we see that there is good agreement i.e. the level spacings of the entanglement spectrum follow the same distribution as the level spacings of the energy spectrum. 
This is true even for $N=18$, where we show data from the entanglement spectrum of a single typical state.\label{DOS_eth}}
\end{figure}

We first study the density of entanglement energies in the orthogonal symmetry class. In Fig.~\ref{DOS_eth} we show density of states for the entanglement spectrum, at several different system sizes. At sizes $N=12,14$, the density of states was obtained by averaging over several thousand states. For $N=16$ and $18$, the data is for only a single state with an energy located in the bulk of the spectrum and obtained using the shift and invert method. This technique was introduced in the context of the ETH-MBL transition in Ref.~\cite{Luitz-PhysRevB.93.060201}.
We see that though the single-state data is noisier, it gives qualitatively the same results as the averaged data. This provides evidence in support of our averaging procedure.  For the GUE case we have obtained data for only $N=12,14$, and the density of states is very similar. Note that in the absence of a reliable numerical library for the shift and invert method for complex matrices, we do not provide any data for $N=18$ when $h_y \neq 0$. 

We next focus on extracting the level spacings of the entanglement spectrum. Fig.~\ref{DOS_eth} shows the distribution of these spacings. For the GOE case, we plot both the level spacings for a single state entanglement spectrum at $N=18$, and that for the average of all states in the shaded region of Fig.~\ref{EnergyDOSLevelStatGOEGUE}a at $N=14$. We also show GUE data taken at $N=14$. 
While the density of states for the entanglement energies barely differs between the orthogonal and unitary classes, the situation is drastically different for the level statistics. Indeed, if the Hamiltonian is in the orthogonal (resp. unitary) class, then the level statistics for the entanglement energies is GOE (resp. GUE). This is a clear example where we see that the level statistics of the Hamiltonian are also encoded in the entanglement spectrum. This agreement of entanglement level statistics in the thermal phase with random matrix theory predictions was also observed in \cite{Chamon}.

The agreement between the entanglement spectrum and the energy spectrum in the thermal phase can be viewed as a consequence of ETH. ETH states that {\it for a small subregion} $A$ in a thermal state, the entanglement Hamiltonian of Eq.~\ref{entanglementHamiltonian} should be proportional to the local Hamiltonian $H_A$ restricted to the subregion $A$, up to an additive constant i.e $\rho_A \approx \frac1Z \exp(- \beta H_A)$ and thus $H_{\rm ent}(\psi, A)=-\log\left(\rho_A\right) \approx \beta H_A + \log Z$  \cite{representativestates}. Here $\beta$ is an inverse temperature related to the energy density in the corresponding eigenstate. So the entanglement spectrum and energy spectrum should have the same level statistics. However, traditional ETH only applies when the subregion $A$ is small (strictly an infinitesimal fraction of the full system), whereas we are considering an entanglement cut whereby $A$ is half the size of the full system. Our results can thus be viewed as evidence in support of a particularly strong form of the ETH (see e.g. \cite{garrisongrover}), which continues to apply even when the subregion is comparable in size to the system itself.

\section{Entanglement spectrum in the localized phase}

We now consider tuning up the amplitude of the random field in Eq.~\ref{ModelHamiltonian}, thereby driving the system into the localized phase, and track the evolution of the entanglement spectrum. The analysis reveals new and hitherto unsuspected entanglement structure in the eigenstates of the many body localized phase, including a residual criticality that persists deep into the localized phase and appears to be a consequence of interactions. 

\subsection{Numerical results}

We recall that the energy spectrum exhibits Poisson level statistics in the localized phase, and is believed to display semi-Poisson level statistics at the critical point \cite{Serbyn}, but once we leave the thermal phase there is no longer any reason to expect the entanglement spectrum and energy spectrum to match. This expectation is clearly borne out by numerics (see Fig.~\ref{EnergyDOSLevelStatPoisson}b), which reveal entanglement spectrum distributions that are definitely not Poisson, even though the energy spectrum is Poisson at the corresponding disorder strength. Moreover we do not observe any noticeable distinctions in the entanglement spectrum of the localized phase depending whether the Hamiltonian is in the orthogonal or unitary symmetry class.

\begin{figure}[htb]
\includegraphics[width=\linewidth]{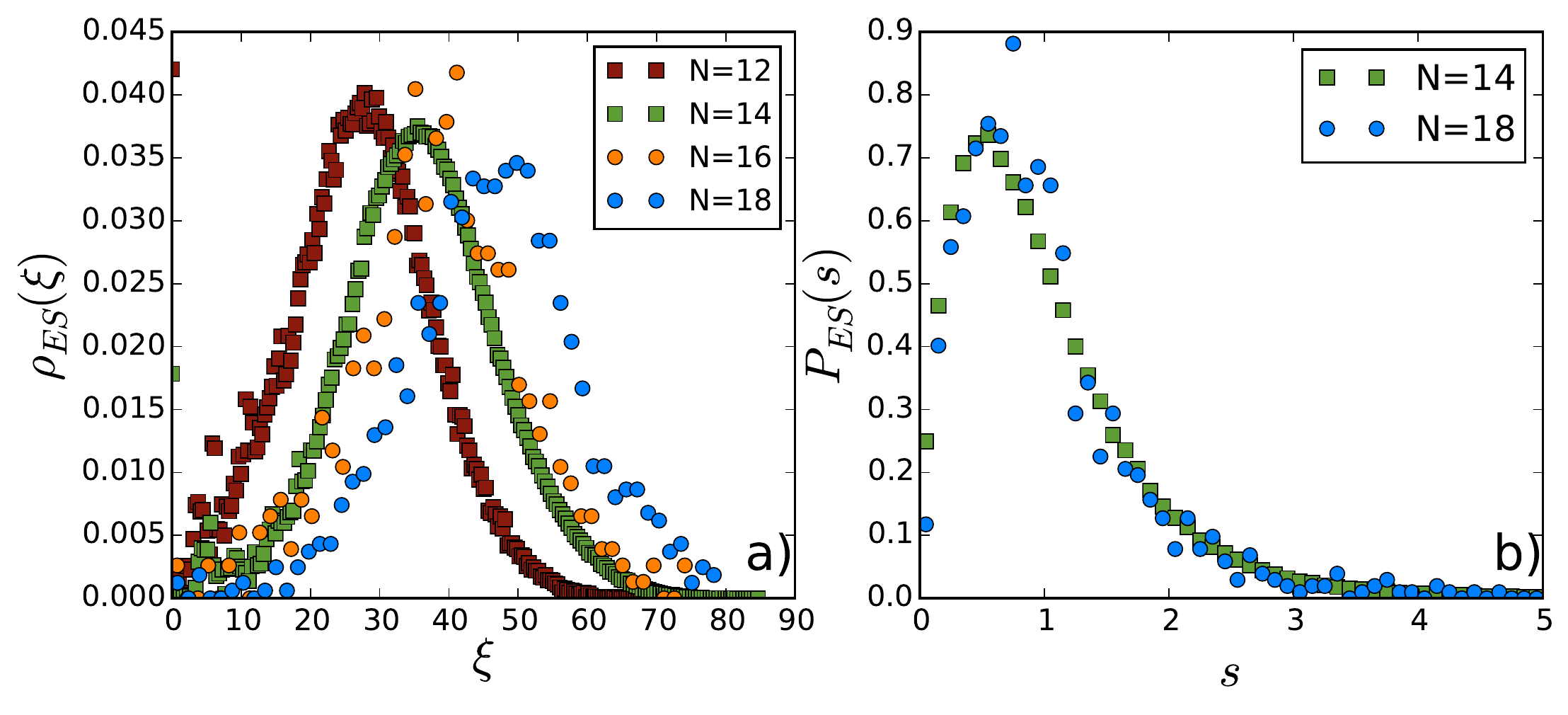}
\caption{ Entanglement spectrum properties with the standard entanglement cut which cuts the system into two equal halves. Data is for $h_x=h_z=12$, and $h_y=0$ (orthogonal class). 
a) Density of states for the entanglement spectrum in the MBL phase  ($h_x=h_z=12$, $h_y=0$).  Note that the density of states has qualitatively the same behavior whether it is extracted from a single entanglement spectrum ($N=16,18$) or an average of many entanglement spectra ($N=12,14$).  
b) Entanglement level splitting distribution averaged over the middle third of the spectrum ($N=14$) of states (red dots), and for a single eigenstate ($N=18$, blue dots).  We note that averaging over a narrow energy window gives the same level statistics as computing the spectrum for a single eigenstate.
\label{DOS_mbl} } 
\end{figure}

 We reexamine the question of whether we can get meaningful results by averaging entanglement spectra, since in the localized phase  there could be dramatic variations in the structure of eigenstates from one eigenstate to the next (which \cite{lbits} referred to as `temperature chaos'). We address this question within the orthogonal class by putting $h_y$ to zero. Fig.~\ref{DOS_mbl} shows the results. It appears that the entanglement level splitting distribution (b) obtained from averaging over a narrow energy window is essentially the same as that obtained from a single eigenstate, so that averaging over eigenstates appears to be an acceptable procedure even deep in the localized phase, if entanglement level statistics are the quantity that is being probed. Thus we conclude that even though the detailed structure of individual eigenstates can change dramatically from one eigenstate to the next in the localized phase, it appears that the statistical properties of the entanglement spectrum do not change, such that microcanonical averages of entanglement spectrum level statistics give essentially the same results as calculations with a single eigenstate. Moreover, the parameters governing the statistical distribution appear to vary only slowly across the energy band such that averaging over a narrow energy window should be a good method for accessing statistical properties of the entanglement spectrum in a numerically efficient manner. 
 
\begin{figure}
\includegraphics[width=0.8\linewidth]{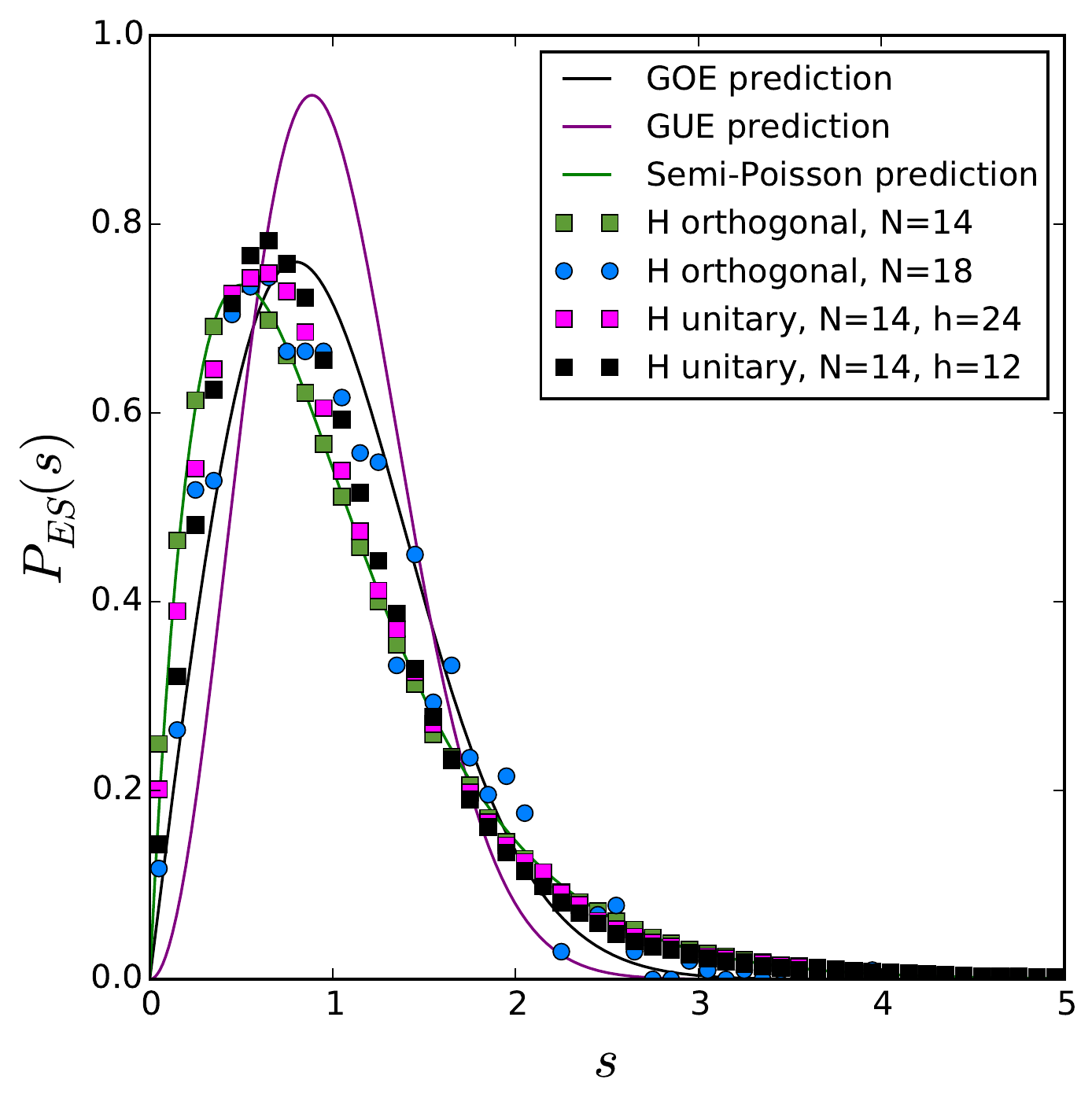}
\caption{A comparison of the ES in the localized phase ($h_x=h_z=12, h_y=0$ for the orthogonal Hamiltonian, $h_x=h_y=h_z=12$ and $24$ for unitary) to the semi-Poisson form $P(s) \sim s \exp(-s)$ (i.e. $\alpha=\gamma=1$). The data in the orthogonal case is the same as that in Fig.~\ref{DOS_mbl}b.  
The $N=18$ data is noisier because we have not averaged over eigenstates. 
\label{spacings_mbl}}
\end{figure}

Having established that averaging over energy eigenstates is an acceptable method of calculating entanglement spectra for small systems, we now turn to a detailed analysis of the entanglement level splittings. We note in Fig.\ref{DOS_mbl}b, the entanglement splittings show {\it level repulsion} but repulsion that is weaker than would be predicted by the Gaussian ensembles. We find that (Fig.~\ref{spacings_mbl}) the level spacing distribution fits beautifully to the {\it semi-Poisson} form $P(s) \sim s \exp(-s)$,\cite{semipoisson_footnote} at least in the orthogonal symmetry class. 
 In the unitary symmetry class the data is qualitatively similar but the fit to the semi-Poisson form is not quite as good. We note that if we go too deep into the localized phase, we run into problems with machine precision as the Schmidt coefficients are extremely small. These machine precision issues are more severe in the unitary class,  are also why we do not show data at larger sizes for the unitary class. 

\begin{figure}[htb]
\includegraphics[width=\linewidth]{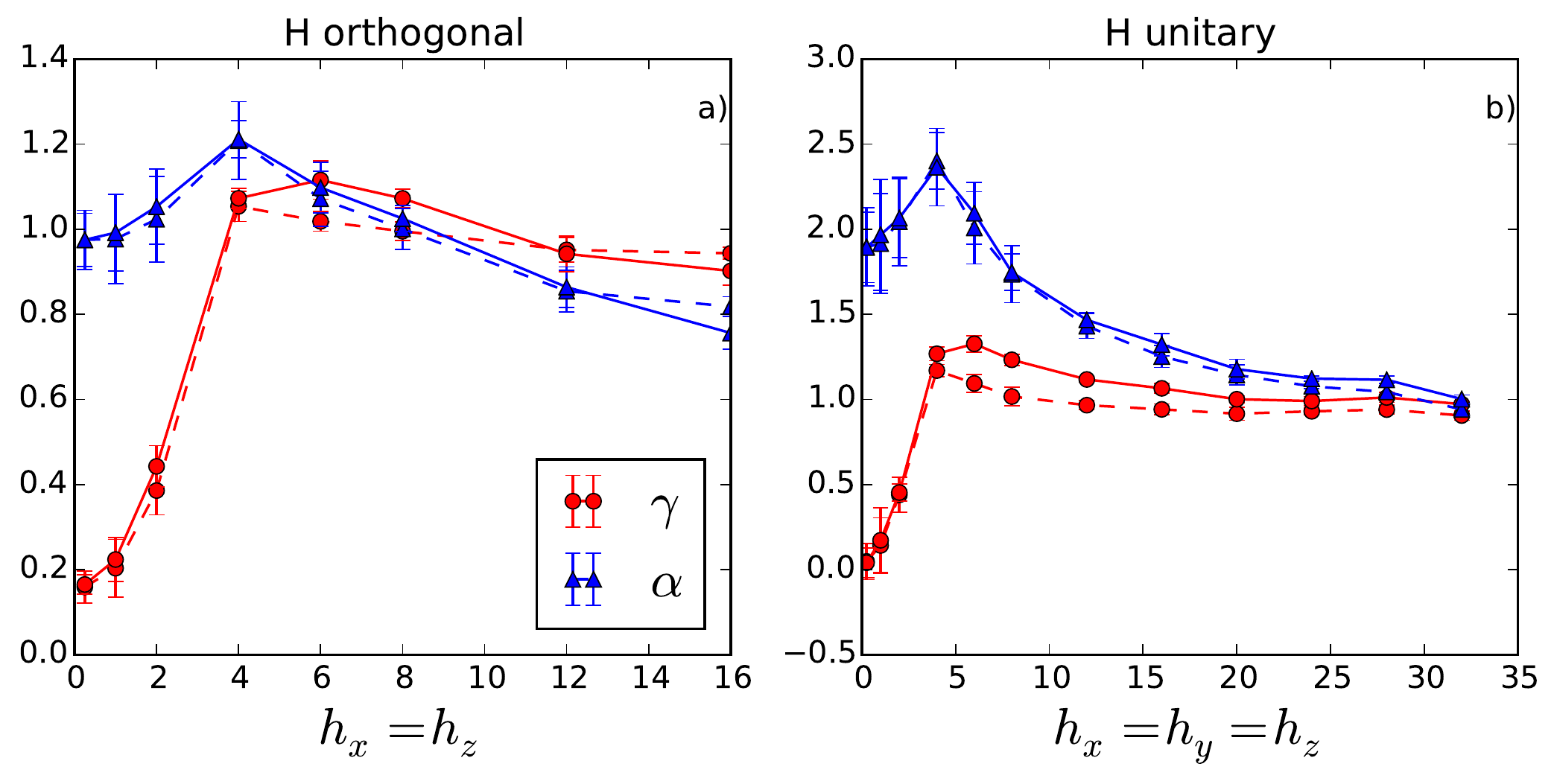}
\caption{
 (a) Evolution of the fitted values of $\gamma$ and $\alpha$ as we tune disorder strength, for an orthogonal Hamiltonian with $h_x=h_z$, $h_y=0$, with $N=14$. The fit is to an ES averaged over states in the middle third of the spectrum. For each exponent we show two curves, each corresponding to a different realization of disorder. There are two sources of error in such measurement: variations between states in the same realization of disorder (ROD), and variations between ROD. The error bars represent the error within an ROD, while the error between ROD can be estimated from the differences between the two curves, and is discussed in more detail in the Appendix.
 \label{ESGOEGammaBeta} 
(b) Same but in the unitary class $h_x=h_y=h_z$.
 \label{ESGUEGammaBeta}}
\end{figure}

For the energy spectrum, the transition from the thermalizing phase to the localized phase has been recently probed in Ref. \onlinecite{Serbyn} by interpolating between the GOE distribution and the Poisson distribution. Indeed, they have proposed to fit the level statistics with the following generalized semi-Poisson distribution
\begin{eqnarray}
P(s)&=&C_1(\gamma,\alpha) s^\alpha e^{- C_2(\gamma,\alpha) s^{2-\gamma}}\label{DistributionGammaBeta}
\end{eqnarray}
where
\begin{eqnarray}
C_2(\gamma,\alpha)=\left(\frac{\Gamma\left(\frac{2+\alpha}{2-\gamma}\right)}{\Gamma\left(\frac{1+\alpha}{2-\gamma}\right)}\right)^{2-\gamma}&\;\;{\rm and}\;\;&C_1(\gamma,\alpha)=\frac{(2-\gamma) C_2^{\frac{1+\alpha}{2-\gamma}}}{\Gamma\left(\frac{1+\alpha}{2-\gamma}\right)}\nonumber\\\label{NormalizationGammaBeta}
\end{eqnarray}
such that the distribution satisfies $<1>=<s>=1$. The Poisson distribution corresponds to $\gamma=1$ and $\alpha=0$, while the GOE (resp. GUE) distribution is obtained for $\gamma=0$ and $\alpha=1$ (resp. $\gamma=0$ and $\alpha=2$). We can also try to see if such a distribution would describe the ES level statistics. We still only consider the case where $h_y=0$. We have fitted the ES level statistics to Eq.~\ref{DistributionGammaBeta}. The resulting values of $\gamma$ and $\alpha$ are depicted in Fig.~\ref{ESGOEGammaBeta}). 
In the localized phase the fit parameters seem quite close to the strict semi-Poisson values $\alpha=\gamma=1$. This is true even for a single realization of disorder. In the Appendix we show data averaged over multiple realizations of disorder, and for a variety of system sizes. Such data are also consistent with $\alpha \approx \gamma \approx 1$ in the localized phase.

\subsection{Criticality in the localized phase ES comes from high entanglement energies and is invisible to entropy measures}

We have found that in the many body localized phase, the ES fits well to a semi-Poisson form, exhibiting level repulsion with small entanglement splittings $s$ being suppressed as $s^{\alpha}$, where $\alpha \approx 1$. Semi-Poisson statistics for the energy spectrum are associated with multifractality and arise at criticality \cite{Serbyn}. The {\it entanglement} spectrum, however, appears to show persistent criticality and multifractality deep into the localized phase. This is a major surprise, since while it is well known that entanglement entropy and Renyi entropies display multifractal statistics {\it at} localization transitions \cite{Gruzberg, Fradkin}, no such features have ever before been observed (or even conjectured to exist, as far as we are aware) deep in the localized phase.

\begin{figure}[htb]
\includegraphics[width=\linewidth]{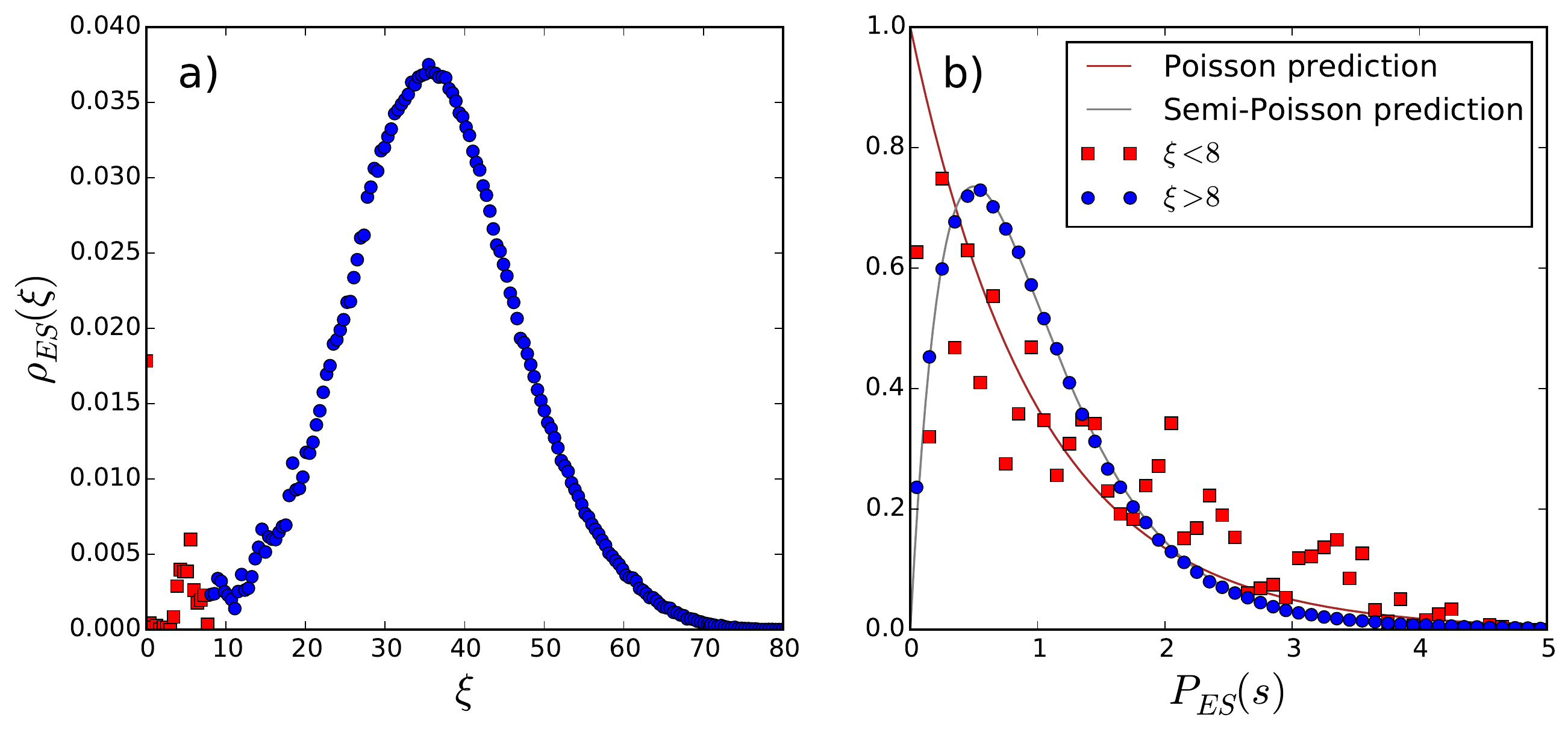}
\caption{a) The density of states data plotted in Fig.\ref{DOS_mbl}a, but only for $N=14$. There are two peaks, once around $\xi=0$ (which contains only $2\%$ of the states), and a much larger one around $\xi=30$. The red points correspond to the `low entanglement energy' part of the spectrum, where we expect non-universal behavior is encoded, while the blue `high entanglement energy' points have the universal distribution of level spacings.
b)  Level spacing distributions for the low and high entanglement energy states. The high entanglement energy states follow a Semi-Poisson distribution with $\alpha=\gamma=1$, while the low entanglement energy states do not. 
\label{separated_spacings}}
\end{figure}

\begin{figure}[htb]
\includegraphics[width=\linewidth]{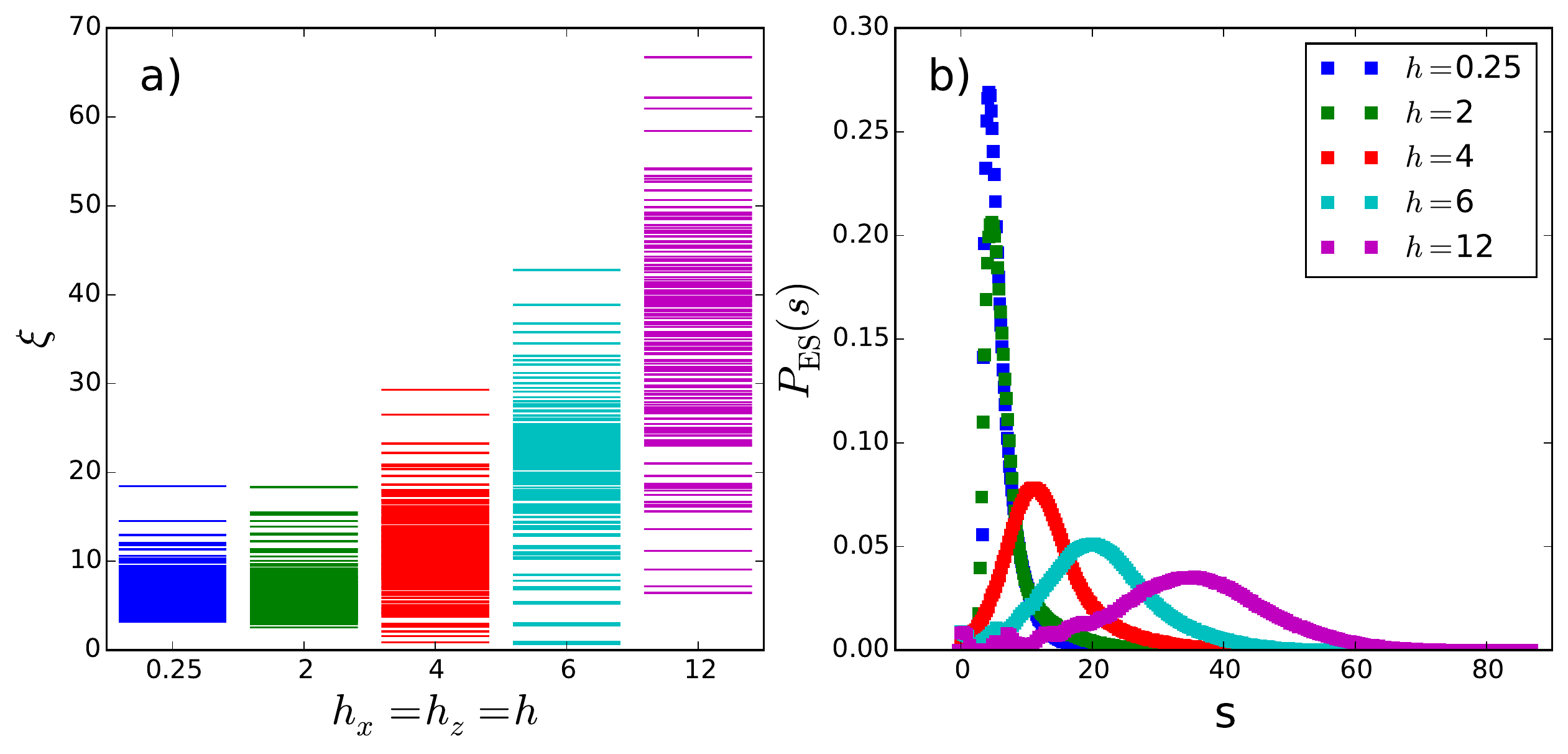}
\caption{(a) Entanglement spectra for a single eigenstate in the middle of the energy spectrum (the 8000th eigenstate in a system with $N=14$), for various values of $h_x=h_z$ and with $h_y=0$. 
(b) Density of states for the same system, averaged over all states in the middle third of the spectrum. We see that most of the states are concentrated in a peak above the entanglement gap, and this peak moves to higher entanglement energies as disorder is increased.
}\label{ESGOECutOff8}
\end{figure}

A partial understanding may be attained by a careful consideration of the density of states in Fig.~\ref{separated_spacings}a for the many-body localized regime. The density of states exhibits a `two peak' structure. There are a small number of states near zero entanglement energy, and the number and location of these states is highly dependent on the realization of disorder. In addition, there are a large number of states above the `entanglement gap'(here concentrated around entanglement energy $ \simeq 30$). This `two peak structure' is illustrated also by Fig.\ref{ESGOECutOff8}. 

An important observation is that the Semi-Poisson statistics that we observe come entirely from the large peak and high entanglement energies.
In the smaller peak below the entanglement gap, there are not enough states to say anything definitive about the statistics of the entanglement levels, but we do not see any signs of level repulsion in these low entanglement energy states. However, as is clear from Figs.~\ref{DOS_mbl} and \ref{ESGOECutOff8}, since the peak above the entanglement gap contains a much larger area, the high entanglement energy states will dominate any measure which averages over all states. The level statistics of the full entanglement spectrum will thus be dominated by the states at high entanglement energy. 
Further evidence is provided in Fig.~\ref{separated_spacings}b, where we plot the level spacing distributions for the high ($\xi>8$) and low ($\xi<8$) entanglement energy (EE) states. The high EE states agree well with the Semi-Poisson distribution, while the low EE states do not show any signs of level repulsion.

Our findings confirm that the signatures of `criticality' in the form of semi-Poisson level statistics come predominantly from states {\it above the entanglement gap} at high entanglement energies. Since most states are located above the entanglement gap, these high entanglement energy states dominate the level statistics of the entanglement spectrum. However, since entanglement entropy and Renyi entropies are dominated by the states at low entanglement energies, the signatures of critical entanglement structure at high entanglement energies will likely be essentially invisible to these entropy measures. They might emerge as correction to the area law but they would be difficult to probe in numerical simulations of small systems. This explains why the remnants of criticality in the entanglement structure of many body localized eigenstates have hitherto been overlooked. Fortunately, these signatures are straightforwardly revealed by an analysis of the full ES. 

This analysis motivates the following understanding for the evolution of the entanglement spectrum across the localization-delocalization transition. Recall that the entanglement entropy (which dominated by the  low entanglement energy states) is volume law in the thermal phase and at the transition \cite{Grover}, but boundary law in the localized phase. Thus, in the thermal phase, there are a large number of states at low entanglement energy, and they obey random matrix statistics with strong level repulsion (governed by the Gaussian ensembles). At the critical point there is still a large number of states at low entanglement energy (a volume law worth), but these show weaker (but non-zero) level repulsion characteristic of multifractality \cite{Fradkin}, and follow a semi-Poisson distribution just like the energy spectrum. At this point, standard entropy measures such as Von Neumann and Renyi entropies also see multifractality. As we move into the delocalized phase, only a small number of states (a boundary law worth) remain at low entanglement energies, and these cease to show level repulsion. Thus, standard entropy measures cease to see any signs of criticality or multifractlity. However, the vast majority of states (a volume law worth) move off to high entanglement energies, and effectively {\it stick} at criticality, continuing to be governed by a semi-Poisson distribution as a signature of multifractality, albeit a signature that is difficult to see in entropy measures, which are dominated by low entanglement energies. 

Thus, signatures of criticality persist in the entanglement structure of eigenstates even deep into the localized phase, but they lie at {\it high entanglement energies} (above the entanglement gap), and flow off to infinite entanglement energy as we go deep into the localized phase and the entanglement gap grows. This residual criticality never disappears, even deep in the localized phase, but it becomes difficult to see in standard entropy measures, which are dominated by low entanglement energies. 

\subsection{Critical entanglement structure as a signature of residual many body entanglement resonances}

We now propose an explanation for the persistent criticality in the entanglement structure of many body localized states. The central idea is that small entanglement gaps correspond to a `resonant' structure in the eigenstates, with different Schmidt states having almost equal weight, and such resonances should be rare in the localized phase, leading to level repulsion of entanglement energies. To make this idea quantitative we employ a {\it gedanken} experiment - consider instead of Eq.\ref{ModelHamiltonian} an otherwise identical system, but with the bond strength at the position of the entanglement cut set to zero i.e. $J_{N/2}=0$ (for a system with periodic boundary conditions one would also have to set $J_0=0$). The Schmidt states $|A:i\rangle \otimes|B:i \rangle$ will now be energy eigenstates, with eigenvalue $E_i$. Now the model of Eq.\ref{ModelHamiltonian} may be recovered by turning $J_{0}$ and $J_{N/2}$ back on. Now the Schmidt states are no longer energy eigenstates. When we construct the energy eigenstates of Eq.\ref{ModelHamiltonian}, the new eigenstates will have some (potentially very small) overlap with all the different Schmidt states. The simplest ansatz is that the overlaps scale as $\exp(-|E-E_{i}|/\zeta)$, where $\zeta$ is some kind of localization length. This ansatz would imply that the entanglement energies are proportional to $|E-E_{i}|$. However, this ansatz cannot be completely correct, since the energies $E_{i}$ will be Poisson distributed, and this logic would imply that the entanglement energies should also be Poisson distributed, which they are not. 

We conjecture that the observed semi-Poisson statistics of the entanglement spectrum can be explained by modifying the above argument to account for {\it many body resonances}. We modify the above ansatz to say that the overlaps scale as $\exp(-|E-\tilde E_{i}|/\zeta)$, where $\tilde E_{i}$ is the energy density of a Schmidt state $|A:i \rangle \otimes |B: i \rangle$ with the full Hamiltonian i.e. $\tilde E_{i} = \langle B:i | \otimes \langle A:i | H |A:i \rangle \otimes |B:i \rangle$, where $H$ is the Hamiltonian from Eq.\ref{ModelHamiltonian}. If the local operator that stitches the system back together (by turning on $J_0$ and $J_{N/2}$) produces a {\it resonance} between two product states $|A:i \rangle \otimes |B:i \rangle$ and $|A:j \rangle \otimes |B:j \rangle$, where a resonance means that the matrix element between these two states is comparable to the energy splitting between them, then this will introduce level repulsion, with the energy splitting $|\tilde E_{i} - \tilde E_{j}| > |E_{i} - E_{j}|$ (note that mean energies of Schmidt states can exhibit level repulsion even though the exact eigenenergies of the Hamiltonian do not). The probability of small splittings $\tilde E_{i} - \tilde E_{j}$ will therefore be suppressed, and the probability of small entanglement energy splittings along with it. 

The probability of small entanglement splittings may be estimated from the probability of small splittings $\omega = \tilde E_{i} - \tilde E_{j}$. In the absence of resonances, these splittings would follow a Poisson distribution. However, a local operator (such as the operator that stitches together the two subregions) will produce resonances, and for an interacting system the density of resonances will scale as $\omega^{-\phi}$, where $0<\phi<1$ (Ref.\cite{response}), with $\phi =1$ being the maximum value consistent with stability of the localized phase (and $\phi \rightarrow 1$ at the critical point).  The distribution of small splittings in the energies of Schmidt states may then be estimated by multiplying the Poisson distribution by a suppression factor $1/\omega^{-\phi}$, where the suppression factor accounts for the level repulsion coming from resonances. The distribution of entanglement splittings will then follow a Poisson distribution multiplied by $\omega^{\phi}$ i.e. will follow $P(s) \sim s^{\phi} \exp(-s)$, where $0<\phi\le1$ and where $\phi \rightarrow 1$ at the transition. This is consistent with our numerical results, which observe a power law suppression of the density of small entanglement energy splittings. Moreover, we observe that the entanglement spectrum (more properly the high entanglement energy part of the entanglement spectrum) {\it saturates} the bound $\phi \approx 1$ (which is expected to apply at criticality), consistent with our picture that the high entanglement energies `stick' at criticality as we move into the localized phase. 

\begin{figure}[htb]
\includegraphics[width=\linewidth]{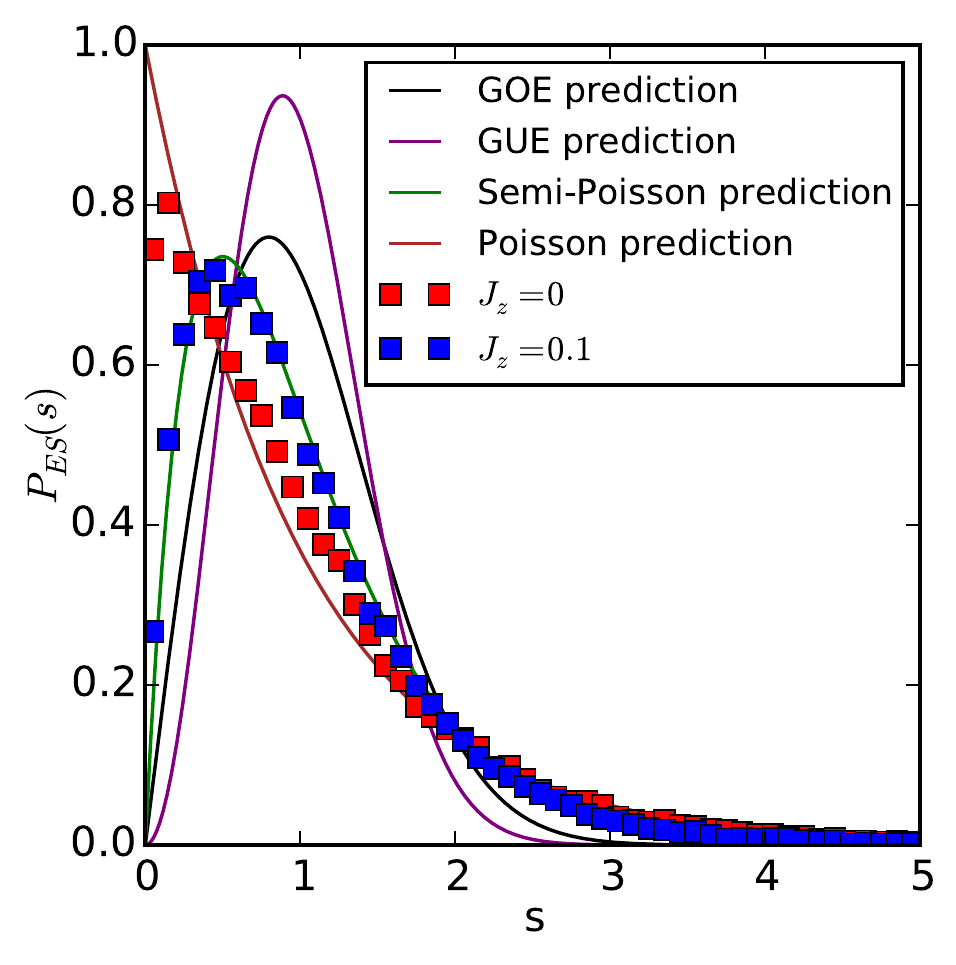}
\caption{Entanglement spacings for a Hamiltonian with $Jx=J_y=1$,$h_x=h_y=0$, $h_z=6$, and varying $J_z$. This system is always in a localized phase, but at $J_z=0$ it is a non-interacting system. We see that in this case (red symbols) the entanglement spectrum does not display the level repulsion observed elsewhere in this work, instead it seems to follow a Poisson distribution. Data was taken for $N=16$ restricted to the $S_z=0$ sector, averaging over the middle third of the energy spectrum. The blue points show that even small interactions return the spectrum to semi-Poisson statistics $\alpha=\gamma=1$.
\label{noninteracting}}
\end{figure}

An interesting test of this conjecture is afforded by non-interacting localized systems. In non-interacting localized systems, many body resonances are absent, and while single particle resonances do exist, their probability scales simply as a logarithmic function of $\omega$ (i.e. $\phi = 0$). We therefore predict, if the above explanation is correct, that non-interacting localized phases should exhibit Poisson level statistics of their ES, with interactions driving a transition to semi-Poisson behavior. That is, we predict that the semi-Poisson form is a diagnostic of the {\it many body} nature of the localized phase, and that a non-interacting localized phase would show Poisson level statistics for the entanglement spectrum. This conjecture may be directly tested by deforming the model of Eq\ref{ModelHamiltonian} by allowing anisotropic interactions $J_x = J_y \neq J_z$ i.e. considering the $XXZ$ spin chain. For $J_z=0$, we have an $XX$ spin chain, which can be Jordan Wigner transformed into a quadratic fermion model. Evaluating the ES of this non-interacting model gives the results in Fig.~\ref{noninteracting}, which shows that at $J_z=0$ the entanglement spectrum loses its level repulsion. 

On turning on the interaction and considering the generic $XXZ$ spin chain we find that the level repulsion is immediately restored. Fig.~\ref{noninteracting} shows that even for small $J_z$, the entanglement spectrum follows semi-Poisson statistics. On the small system sizes studied here we can find a transition from the Poisson to semi-Poisson behavior at small $J_z$. The critical $J_z$ shrink rapidly to zero as system size is increased, so we expect that in the thermodynamic limit the Poisson distribution is unstable even to small interactions. This conclusively demonstrates that the level repulsion observed in the entanglement spectrum is intimately tied to the {\it interacting} nature of the localized phase, and reveals a new feature in the entanglement structure of eigenstates that distinguishes single particle and many body localized states. It also provides supporting evidence for our conjectured explanation of the level repulsion as arising due to many body resonances.  

\subsection{Summary of results on one dimensional Hamiltonian systems}
We now summarize the key results from our study of the ES in one dimensional Hamiltonian systems. In the thermal phase the level statistics of the ES follow the predictions of random matrix theory (for the same ensemble as the energy spectrum) in accordance with a strong version of the ETH. In the localized phase, the level statistics of the ES follow a semi-Poisson distribution. In both the thermal and the localized phases, the statistical properties of the ES do not vary much from one eigenstate to the next, so that averaging over eigenstates in a narrow energy window provides essentially identical results (with lower statistical error) to calculations for a single eigenstate. 

The semi-Poisson level statistics of the ES in the localized phase are indicative of criticality in the entanglement structure, which surprisingly persists deep into the localized phase. However, these residual signatures of criticality are difficult to see in standard measures such as entanglement entropy or Renyi entropies, since they are absent in the low entanglement energy states which dominate these entropy measures, and are carried instead by the parts of the entanglement spectrum at high entanglement energy. Apparently the ES evolves across the transition as follows: at the critical point, all entanglement energies follow semi-Poisson statistics, just like the energy levels. As we move into the localized phase, the low entanglement energy states (which dominate the entropy measures) cease to show level repulsion, but the high entanglement energy states `stick' at criticality and continue to follow the semi-Poisson form. As we move deep into the localized phase, these states move to ever higher entanglement energy (i.e. have ever less weight in the many body wavefunctions), but they continue to show signs of residual criticality up to the largest disorder strengths that we can probe (up to $h\approx 30$). We have also linked this residual criticality to interactions (it is absent in non-interacting localized phases), and have advanced an explanation of the apparent level repulsion in the ES in terms of many body resonances. However, all our results thus far have been restricted to one dimensional systems.

\section{Two dimensional transverse field Ising model}

The vast majority of existing work on many body localization and thermalization has focused on one dimensional systems, with higher dimensional systems representing an important open problem. In this section, we investigate the evolution of the ES across a two dimensional localization delocalization transition, using numerical exact diagonalization. We find results which are qualitatively identical to one dimensional systems, suggesting that the entanglement structure in the thermal and localized phases is insensitive to spatial dimension. 

\begin{figure}[htb]
\includegraphics[width=0.45\linewidth]{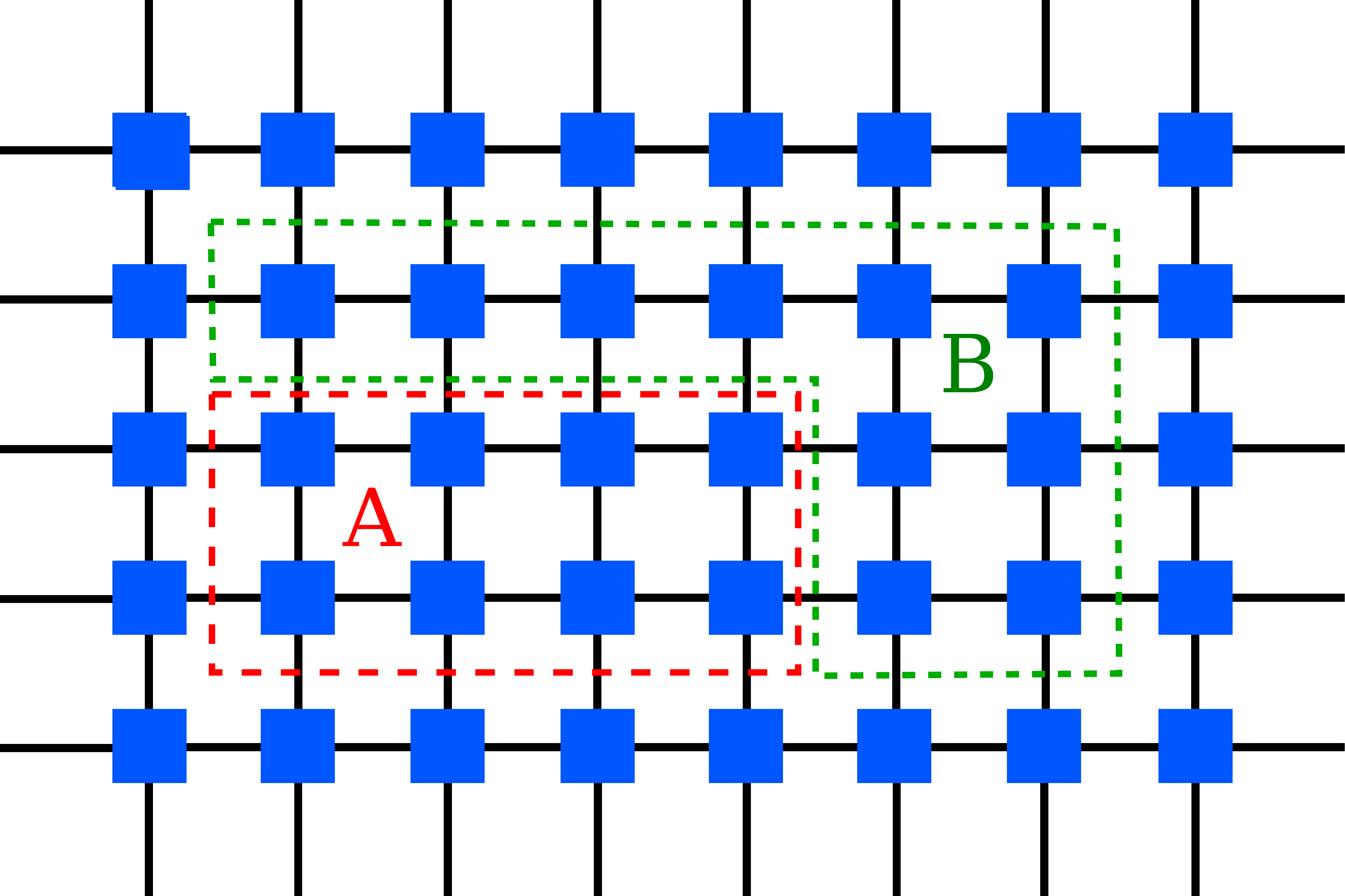}
\caption{The two dimensional square lattice used for the transverse field Ising model. Here we depict the case that we have considered, namely $N=18$ spin-$1/2$ on square torus of $N_x=6$ sites in one direction and $N_y=3$ in the perpendicular direction. When we perform the ES, the system is cut into a part $A$ (red dotted region) of $N_A=8=4\times2$ sites and a part $B$ (green dotted region) made of the $10$ remaining sites. Note that the cut breaks both translation symmetries and the inversion symmetry.}\label{Lattice2DTFIM}
\end{figure}

\begin{figure}[htb]
\includegraphics[width=\linewidth]{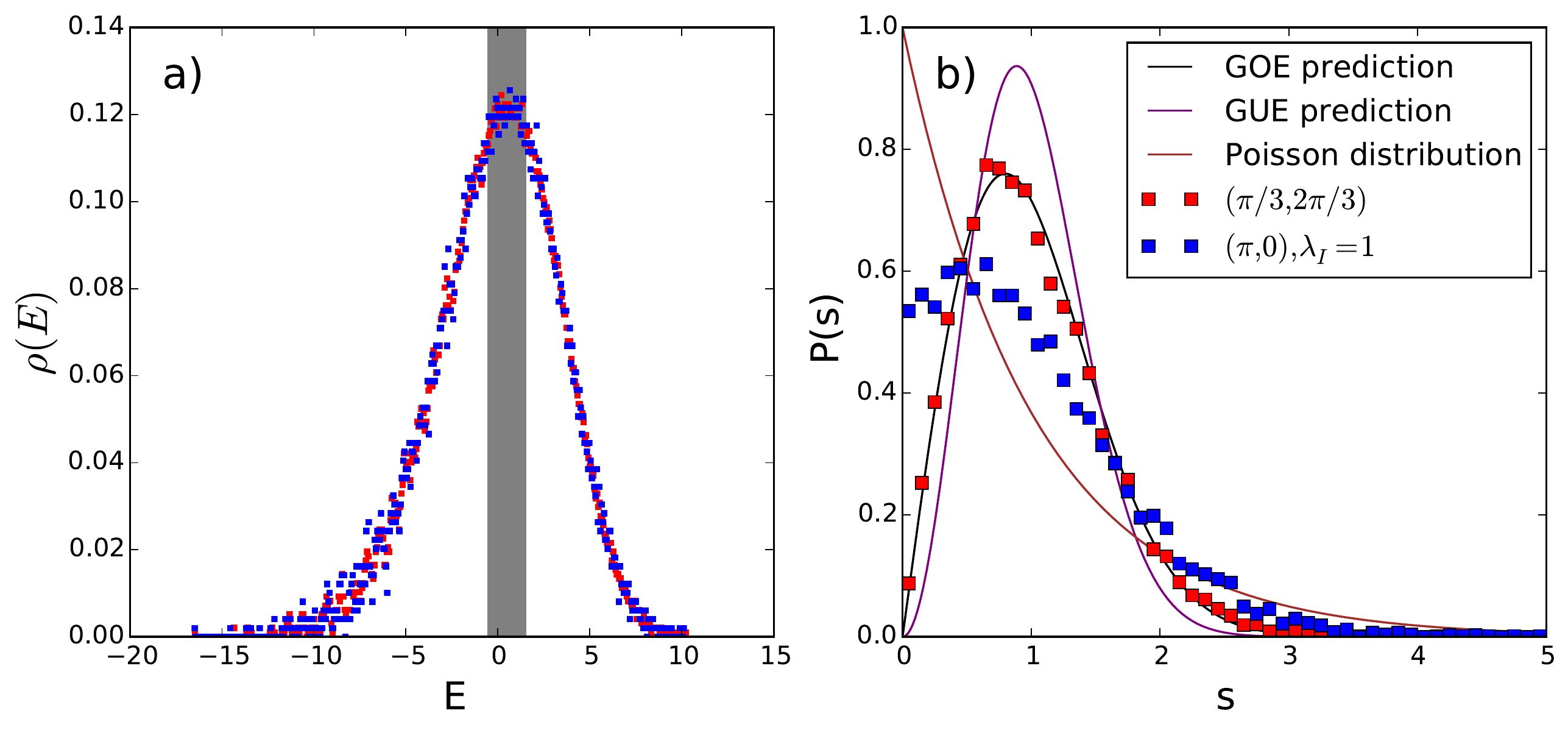}
\caption{a) Energy spectrum density of states and b) energy level statistics for the system of $N=18$ spins depicted in Fig.~\ref{Lattice2DTFIM}. Here we have used $h_x=h_z=-J=1$.We look at a typical momentum sector not invariant under the inversion symmetry ($(\pi/3,2\pi/3)$ black dots) and a typical momentum sector invariant under the inversion symmetry ($(\pi,0)$). In the right panel, we clearly see that level statistics is given by the GOE for the ($(\pi/3,2\pi/3)$ sector while it is somehow closer to the Poisson distribution for $(\pi,0)$ (the same is true in other inversion symmetric sectors). We also show in the left panel the region of the spectrum (shaded area) that we will consider for the entanglement spectrum analysis, setting $i_{\rm min}=6000 $ and  $i_{\rm max}=9500$ for $(\pi/3,2\pi/3)$ ($\simeq 24\%$ of all states in this sector) and $i_{\rm min}=3000 $ and  $i_{\rm max}=4500$ for $(\pi,0)$ and $\lambda_{\rm I}=\pm 1$($\simeq 20\%$ of all states in this sector). }\label{EnergyDOSLevelStat2DTFIM}
\end{figure}

We make use of the model from Ref.~\onlinecite{Mondaini-2015arXiv151204947M}, which has recently provided convincing evidence of the ETH for the two dimensional transverse field Ising model using exact diagonalizations on small clusters. We consider $N$ spin-$1/2$ on a square lattice of size $N_x \times N_y$ (see Fig.~\ref{Lattice2DTFIM}) with periodic boundary conditions. The transverse field Ising Hamiltonian is 
\begin{widetext}
\begin{eqnarray}
H_{\rm 2DTFIM}&=& J \sum_{<{\mathbf{i}},{\mathbf{j}}>} S^{z}_{\mathbf{i}} S^{z}_{\mathbf{j}}+h_x\sum_{\mathbf{i}}  S^{x}_{\mathbf{i}}+h_z\sum_{\mathbf{i}}  S^{z}_{\mathbf{i}}
+h_{z_{\rm random}}\sum_i c_i S_i^z
\label{Hamiltonian2DTFIM}
\end{eqnarray}
\end{widetext}
where $<{\mathbf{i}},{\mathbf{j}}>$ denotes the pairs of neighboring sites. We first look at the disorder free case ($h_{z_{\rm random}}=0$). This is informative in two dimensions since (unlike in $d=1$) unless we look at some finely tuned values of $h_x$ and $h_z$, the model is not integrable. In the following and without any loss of generality, we will focus on the ferromagnetic case $J<0$. For the numerical calculations we have used both the translation symmetries along the two directions $x$ and $y$ and the inversion symmetry. The momentum is denoted $\left(k_x,k_y\right)$ and the inversion parity $\lambda_{\rm I}=\pm 1$. As observed in Ref.~\onlinecite{Mondaini-2015arXiv151204947M}, the level statistics of the energy spectrum differ if we look at one of the inversion symmetric momentum sectors (i.e. $(0,0)$, $(0,\pi)$, $(\pi,0)$ or $(\pi,\pi)$) or any other momentum sectors. In sectors without inversion symmetry the distribution follows the predictions of random matrix theory for the GOE, whereas in sectors with inversion symmetry the distribution is close to (but not quite) Poisson.  This property is illustrated by Fig.~\ref{EnergyDOSLevelStat2DTFIM} for a system of $N_x \times N_y = 6 \times 3$ spins.
 
\begin{figure}[htb]
\includegraphics[width=\linewidth]{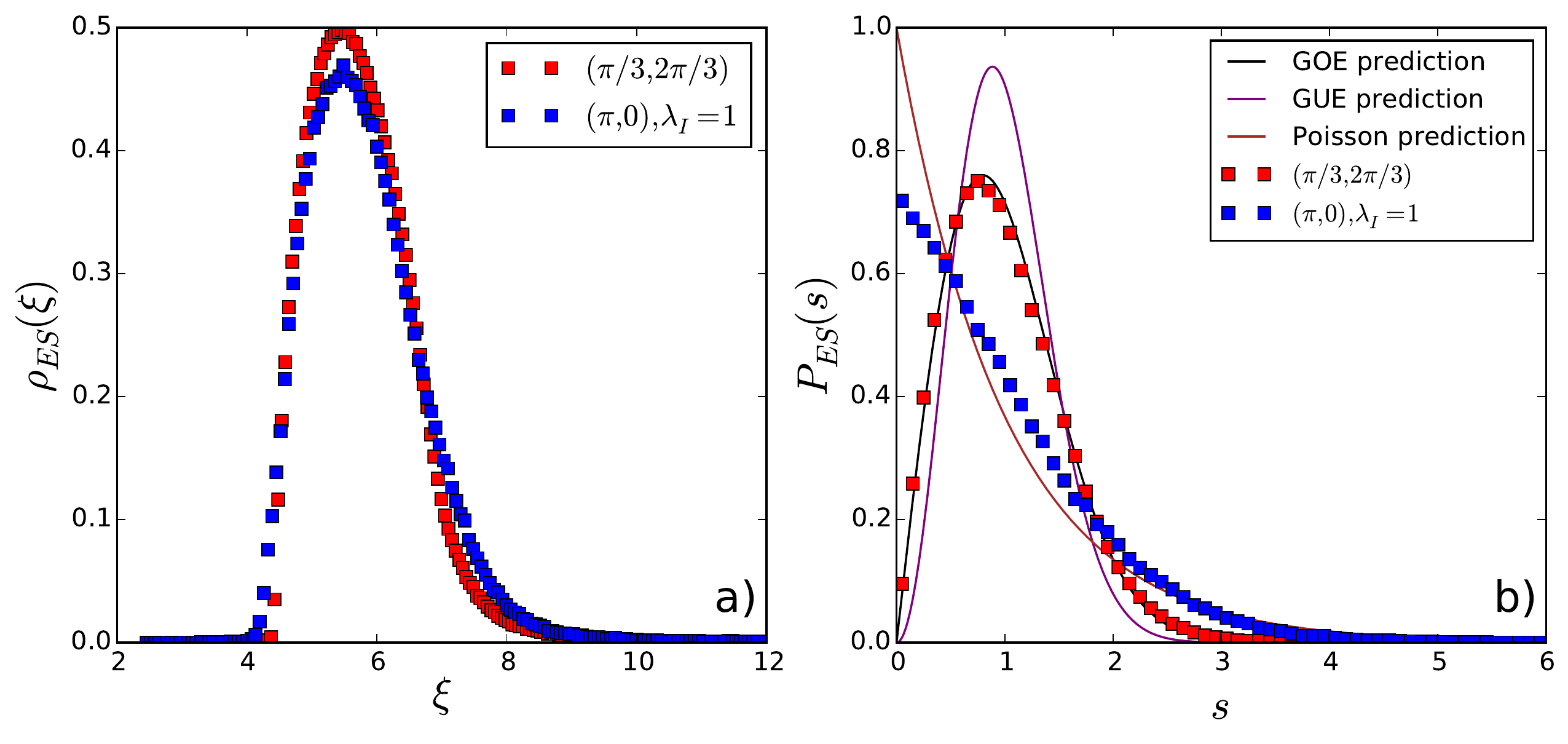}
\caption{Density of states (a) and level statistics (b) for the entanglement spectra of the system with $N=18$ spins. The system geometry and the bipartition are shown in Fig.~\ref{Lattice2DTFIM}. Here we consider to the  Hamiltonians of Eq.\ref{Hamiltonian2DTFIM} with $h_x=h_z=-J=1$. We look the same momentum sectors than those of  Fig.~\ref{DOSLevelStat2DTFIMEntSpectrum}. The average over the entanglement spectra is done over the $3500$ states ($1500$ for each sector of  $(\pi,0)$) located in the bulk of the spectrum as shown in Fig.~\ref{EnergyDOSLevelStat2DTFIM}a. }\label{DOSLevelStat2DTFIMEntSpectrum}
\end{figure}

We can now look at the properties of the entanglement spectrum for this model. For simplicity, we choose to cut our system in such a way that the ES does not inherit any of the quantum numbers of the original system. This can be easily achieved by taking a rectangular domain that break both the translation symmetries and the inversion symmetry as shown in Fig.~\ref{Lattice2DTFIM}. We use the same setup as the one discussed in Sec.~\ref{Sec:ETHES}, i.e. we average over a large number of ES corresponding to states in the bulk of the spectrum. The average level statistics for the entanglement spectrum is given in Fig.~\ref{DOSLevelStat2DTFIMEntSpectrum}. We observe the remarkable high similarity between the energy spectrum level statistics and  the entanglement spectrum level statistics.

We now want to study the transition to the localized phase, which we will accomplish by tuning $h_{z_{\rm random}}$. Since no-one has studied this transition before, we begin by locating the transition by looking at the distributions of the energy eigenstates.
To numerically probe this transition, we can compute the ratio of adjacent energy gaps. For a sorted spectrum $\{\lambda_n;\lambda_n \leq \lambda_{n+1}\}$, the ratio of adjacent gaps is defined as
\begin{eqnarray}
r_n&=&\frac{{\rm min}\left(\lambda_n - \lambda_{n-1}, \lambda_{n+1} - \lambda_{n}\right)}{{\rm max}\left(\lambda_n - \lambda_{n-1}, \lambda_{n+1} - \lambda_{n}\right)}\label{RatioAdjacentGap}
\end{eqnarray}
We incorporate the full energy spectrum in the calculation of $r$. The average ratio $r$ of adjacent gaps is $r \simeq 0.530$ for GOE\cite{dAlessio}, $r \simeq 0.60$ for GUE \cite{dAlessio} and $r \simeq 0.386$ for a Poisson spectrum \cite{Pal}. 

\begin{figure}[htb]
\includegraphics[width=\linewidth]{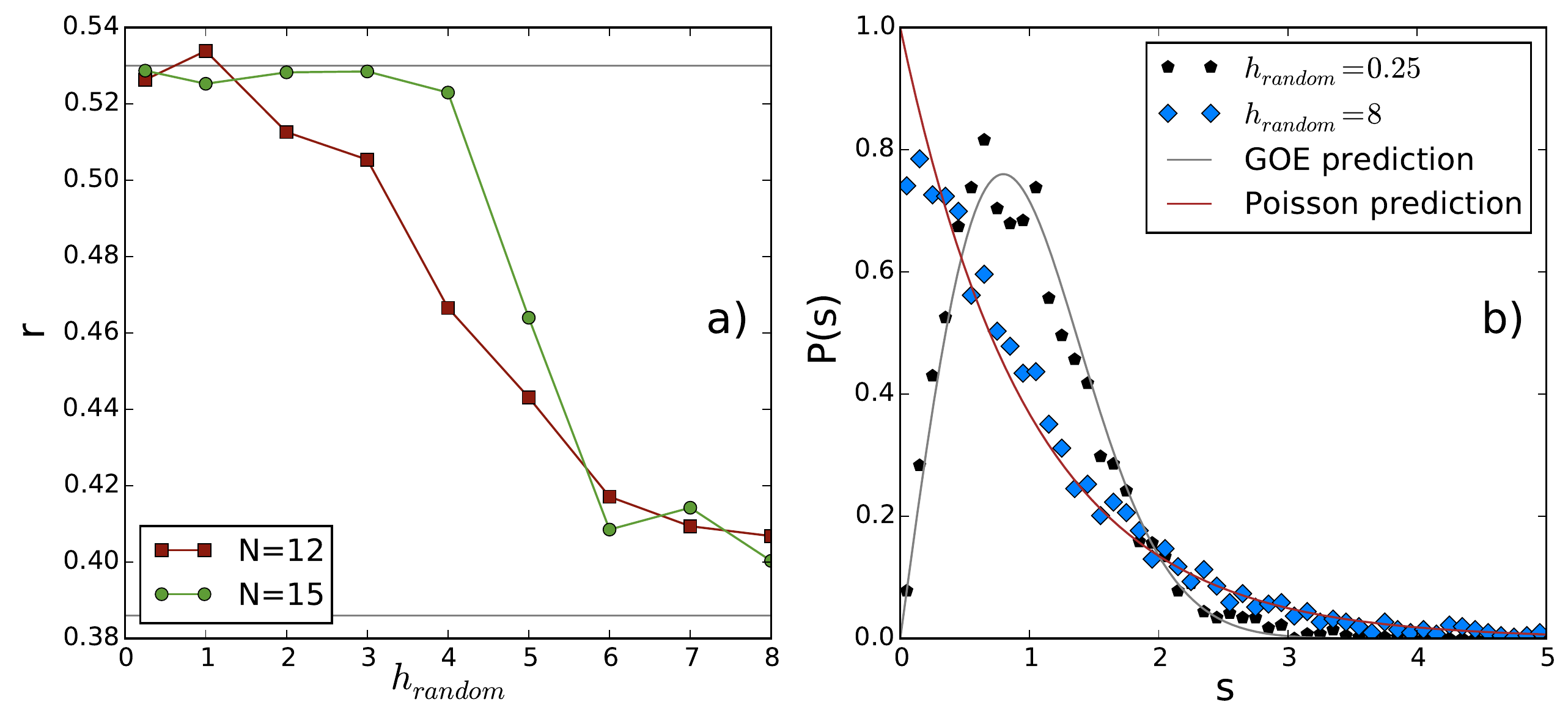}
\caption{(a) The $r$ level spacing parameter for the energy eigenstates for the $2D$ model. The grey lines are a guide for the eye showing the predicted values for the Poisson ($r\approx 0.386$ \cite{Pal}) and GOE ($r\approx 0.530$ \cite{dAlessio}) cases. We see a transition from GOE to Poisson. For the $N=15$ data, only the $200$ states near the middle of the energy spectrum were used, while the $N=12$ data includes all states. (b) Energy level spacing distributions for a system with $N=12$ at representative points in the ETH and MBL phases, which follow the correct distributions. }\label{2D_energy}
\end{figure}

Fig.~\ref{2D_energy} a) shows the  $r$ for systems of $N=4\times 3=12$ and $N=5\times3=15$ states. We see that it starts at the $GOE$ value of $0.52$, and as $h_{\rm random}$ is increased it drops to the Poisson value of $0.396$ at around $h_{\rm random}=5$. 
Fig \ref{2D_energy} b) shows distributions of the spacings of the energy levels, which agree with the GOE and Poisson distributions in the appropriate phases.

Now we can ask how the entanglement spectrum behaves for the same system. When making entanglement cuts in 2D, we have some freedom of choice as to where to make the cut. At $N=12=4\times 3$, we cut a $3\times2$ region out, similar to Fig.\ref{Lattice2DTFIM} but for the different size. Similarly for $N=15=5\times 3$ we cut a $4\times2$ region. Such cuts are advantageous in the absence of disorder since they break all symmetries, but in the disordered case we are free to choose other cuts, since the disorder breaks symmetries anyway, and for $N=16=4\times4$ we choose a $4\times2$ cut. 

\begin{figure}
\includegraphics[width=\linewidth]{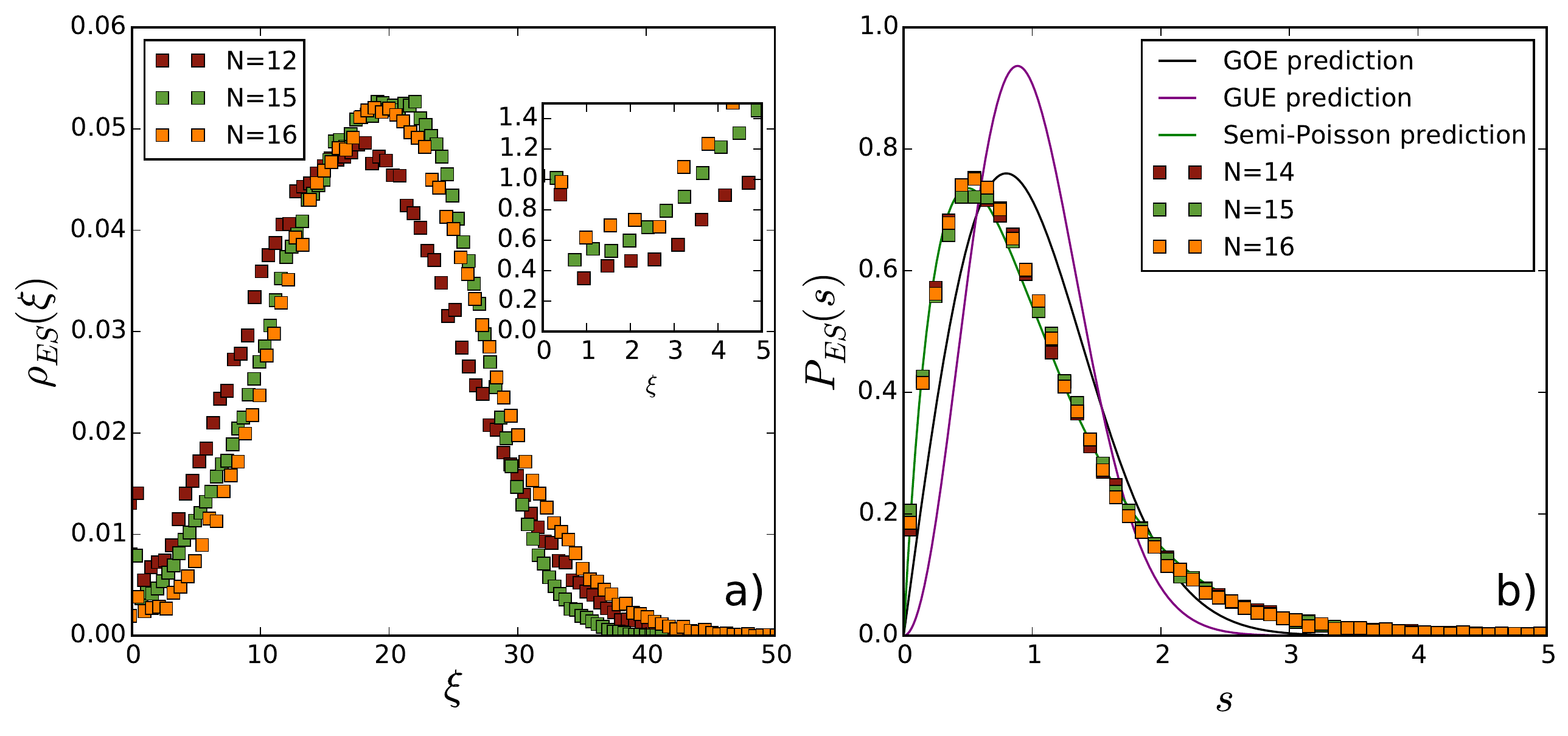}
\caption{ (a)Entanglement density of states for the 2D case. As in the $1D$ case, we see a small peak at $\xi=0$ as well as a larger peak at higher entanglement entropies. The data is normalized such that each curve has an equal area. This hides the behavior at large $N$ and small $\xi$, so in the inset we plot data normalized such that each curves area is proportional to the number of states, and the small $\xi$ peak is visible. Data is taken for $h_{\rm random}=8$.
(b)Entanglement level spacings for the 2D case, which follows a strict semi-Poisson distribution $\alpha=\gamma=1$. Data is taken for $h_{\rm random}=8$. We see that the data matches well the expected distribution shown by the solid green line. }\label{2D_DOS}
\end{figure}

First we show entanglement density of states for a variety of system sizes in Fig. \ref{2D_DOS}. The figure is qualitatively very similar to Fig.\ref{DOS_mbl}, with a peak at very small values but the bulk of the states in another peak at larger values. Once again the data is normalized so that the area is proportional to the number of entanglement states. 

We can also look at entanglement level spacings in the MBL phase, which are shown in Fig.~\ref{2D_DOS}, and follow a semi-Poisson distribution, similar to the 1D case.
In the Appendix we fit the entanglement statistics to Eq~(\ref{DistributionGammaBeta}) and find that once again the exponents are quite close to the strict semi-Poisson values $\alpha = 1 = \gamma$, even though the energy spectrum is Poisson $\alpha = 0$. Once again we do not see any sign of level repulsion in the states at low entanglement energy - but the vast majority of the states lie at high entanglement energy and follow the semi-Poisson form. 

To summarize, the statistical properties of the ES in both the thermal and the localized phase appear substantially similar in this two dimensional model to the results we obtained earlier in one dimension. We conclude therefore that the entanglement structure of the eigenstates does not appear to depend strongly on spatial dimensionality.

\section{Entanglement spectrum in Floquet systems}
Thus far we have focused on the ES in the thermal and localized phases of {\it Hamiltonian} systems, in spatial dimensions $d=1,2$. We now consider periodically driven {\it Floquet} systems, which constitute a different universality class due to the absence of energy conservation. We find that the entanglement spectrum in the localized and thermal phases both is substantially similar to that observed in Hamiltonian systems. In the thermal phase this is something of a surprise since Floquet systems thermalize to infinite temperatures, such that the ETH does not relate the entanglement Hamiltonian to the real Hamiltonian. Nevertheless we find that the entanglement spectrum in Floquet systems follows predictions of random matrix theory, which appears to be more generally applicable than ETH. We also find that the choice of random matrix ensemble can depend on the choice of origin of time for the Floquet operator.

We focus on a simple model based on a chain of $N$ spins-$1/2$ with periodic boundary conditions. We are using a two bang approach for the time evolution
\begin{equation}
U(\tau) = \exp(-i H_1 \tau/2) \exp(-i H_2 \tau/2).\label{TwoBangEvolution}
\end{equation}

We consider the situation where both $H_1$ and $H_2$ are integrable
\begin{eqnarray}
H_1&=&\sum_{i=1}^{N} S^{x}_i S^{x}_{i + 1}+ h c_{x,i} S^{x}_i,\label{H1Integrable}\\
H_2&=&\sum_{i=1}^{N} S^{z}_i S^{z}_{i + 1}+ h c_{z,i} S^{z}_i.\label{H2Integrable}
\end{eqnarray}

\begin{figure}[htb]
\includegraphics[width=\linewidth]{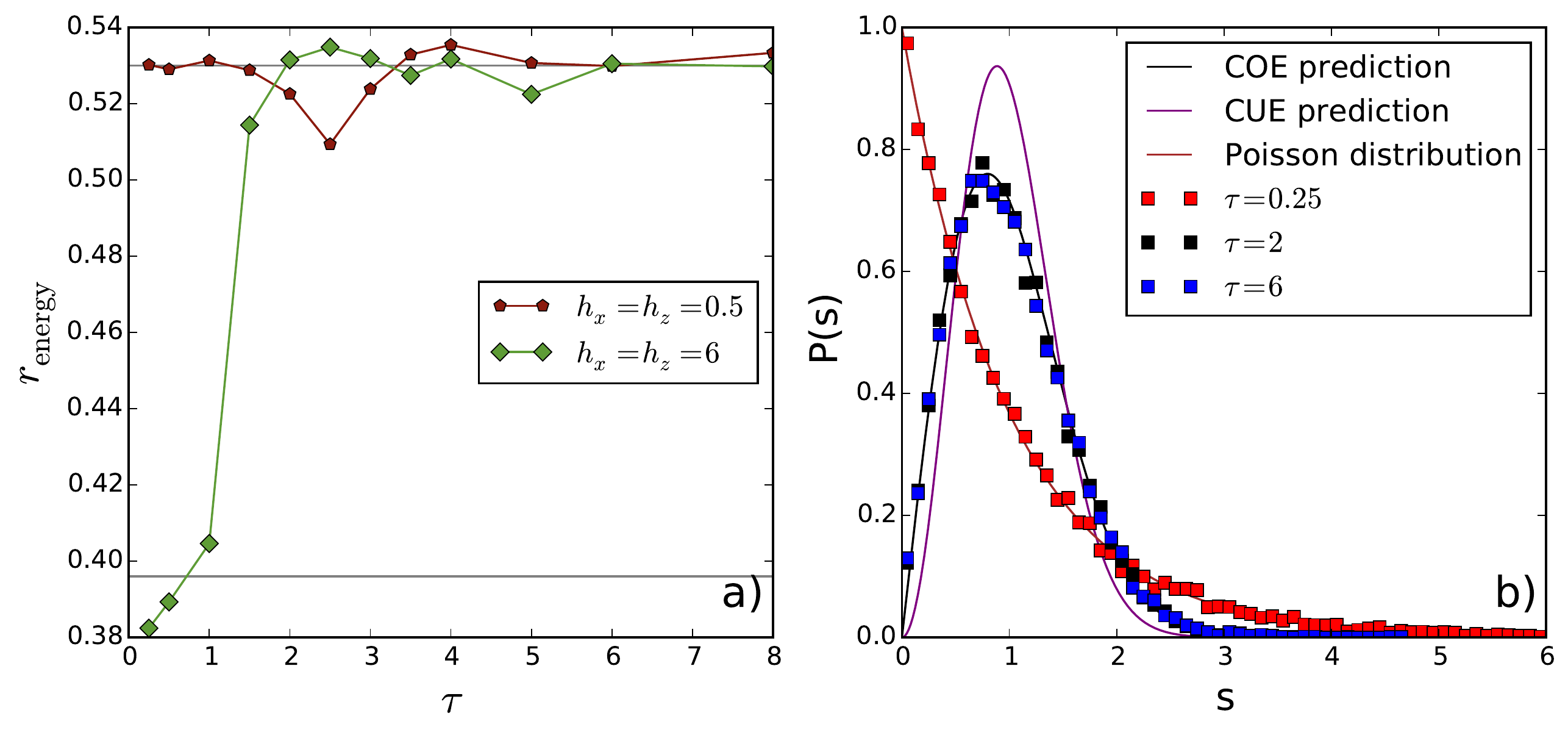}
\caption{a) Ratio of adjacent gaps $r_{\rm energy}$ for the Floquet energies of the two bang-models where $H_1$ and $H_2$ are given by Eqs.~\ref{H1Integrable} and \ref{H2Integrable}. The calculations were done for $N=14$ spins for a single realization of disorder. The Floquet energy data shows the same behavior of that in Ref.~\onlinecite{Regnault}. Grey lines show the predicted values for the Poisson ($r\approx 0.386$ \cite{Pal}) and COE ($r\approx 0.530$ \cite{dAlessio}) cases.
b) Level statistics (right) for a system of $N=14$ spins for $h=6$. Even without any average over disorder, we clearly see the good agreement with either the Poisson distribution for $\tau=0.25$ or the COE for $\tau=2$ and $\tau=6$.}\label{EnergyDOSLevelStatIntegrableCOE}
\label{IntegrableCOE}
\end{figure}

The level statistics of the Floquet Hamiltonian was discussed in Ref.~\onlinecite{Regnault}, for completeness we briefly summarize it here.
Since the Floquet operator is unitary, we are looking at the phases of its eigenvalues, which we call here the `Floquet energies'.
At small $h$ ($\lesssim 2$) the Floquet energy statistics follow a circular orthogonal ensemble (COE), (which has the same spacing distribution as the GOE distributions studied in this work) at both small and large $\tau$. There is however a `dip' in the parameter $r$ when the Floquet bandwidth is similar to the drive frequency. At large $h$ ($\gtrsim 2$) the Floquet energy distribution is localized (obeys a Poisson distribution) at small $\tau$ and thermal (obeys a COE distribution) at large $\tau$.
These properties can be seen in Fig.~\ref{IntegrableCOE}a, where we show the parameter $r$ as a function of $\tau$ for $h$ values representative of the thermalized and localized phases. Unlike the data in Ref.~\onlinecite{Regnault}, here we use only one realization of disorder, and we use a system size large enough ($N=14$) to get adequate statistics despite this. 
In Fig.~\ref{IntegrableCOE}b we show the density of states and spacing distribution for the Floquet energy levels at $h=6$, which further supports the conclusions above.

\begin{figure}[htb]
\includegraphics[width=\linewidth]{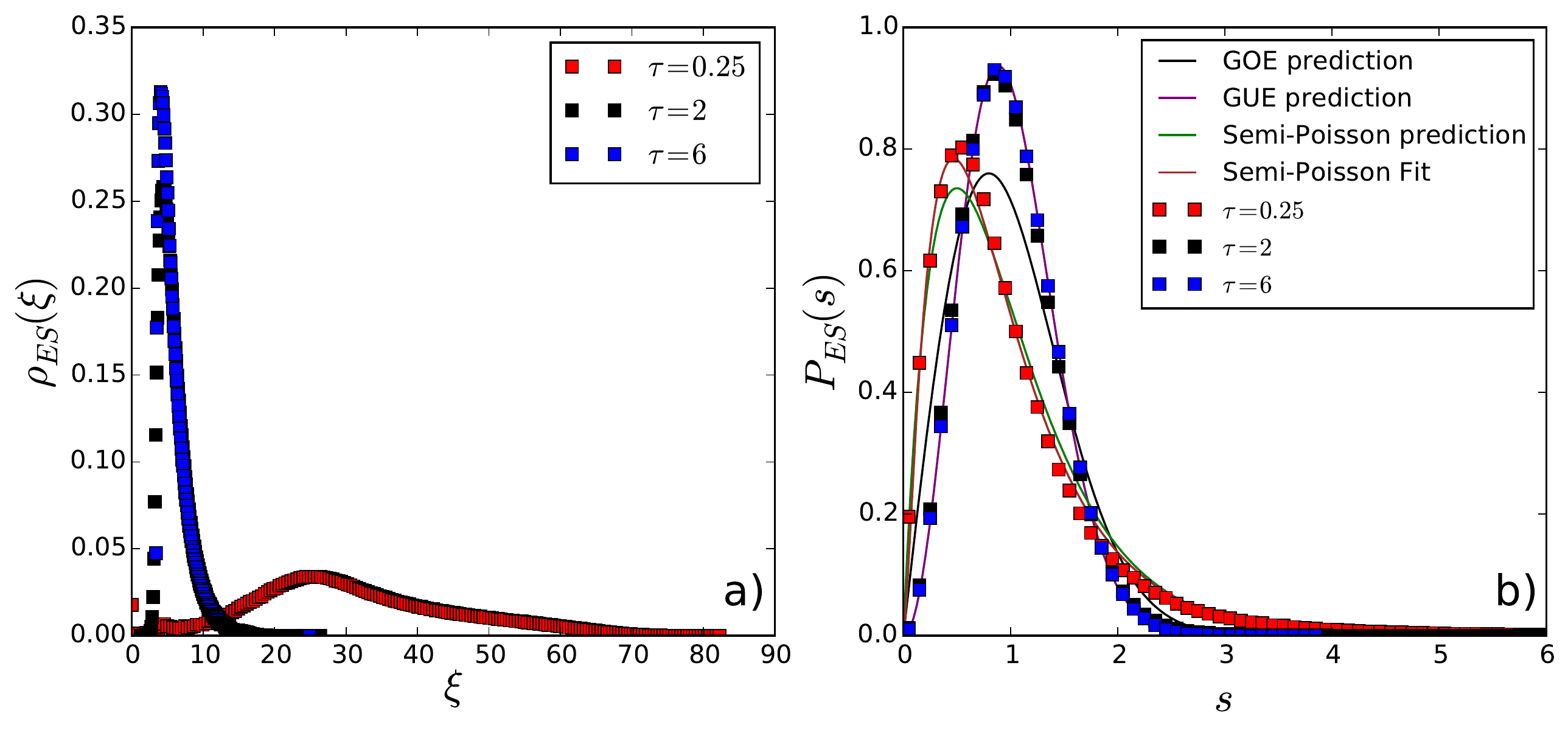}
\caption{Density of states (a) and level statistics (b) for the entanglement spectra of a system with $N=14$ spins keeping $N_A=7$ spins and $h=6$. Here we consider evolution times $\tau=0.25$ (blacks dots), $2$ (blue dots) and $6$ (red dots). The average over the entanglement spectra is done over the all the $4096$ states. The origin of time is chosen according to Eq.~\ref{TwoBangEvolution}. Considering only a fraction of the states, does not change the picture. The level statistics in the thermal phase are well described by the GUE. In the localized phase ($\tau=0.25$), there is a good fit to the strict semi-Poisson form $\alpha=\gamma=1$ (green line). We have also fitted the level spacing distribution to the interpolating distribution considered in Ref.~\cite{Serbyn}. The fitting parameters are $\gamma\simeq 1.68(6)$ and $\alpha=1.31(2)$ (red line). [Errors bars are from the fits alone and do not include averaging over disorder or finite size effects]}\label{Floquet_DOS} 
\end{figure}

We now discuss the entanglement spectrum. Fig~\ref{Floquet_DOS}b shows our results for the entanglement spacings of a Floquet system. We see that as in the undriven case, in the localized phase the entanglement statistics seem to follow a semi-Poisson distribution. The story in the thermalizing phase however is far richer. 

When Floquet systems thermalize, they do so to infinite temperature \cite{Nandkishore-2015}. The ETH therefore predicts that in a thermalizing Floquet phase, the reduced density matrix will be proportional to the unit operator, and the entanglement Hamiltonian will be {\it zero}, such that the entanglement spectrum will be totally degenerate. In practice, however, (and certainly for a finite size system), we expect there will be corrections to ETH and these corrections will {\it dominate} the entanglement spectrum. The entanglement spectrum thus provides a nice diagnostic for investigating {\it corrections} to ETH, since in the Floquet thermal phase these corrections to ETH provide the leading contribution to the entanglement spectrum. We observe indeed that the entanglement spectrum in the Floquet phase fits well to predictions of random matrix theory, indicating that random matrix theory is a {\it more} robust descriptor of thermalization than ETH i.e. even the corrections to ETH follow random matrix theory. We also  highlight a surprising feature of the entanglement spectrum in the thermal phase : the entanglement spacings follow a GUE distribution, even though the energy spacings are GOE i.e. the entanglement spectrum in the thermal phase is governed by a different random matrix ensemble than the energy spectrum. 

\begin{figure}[htb]
\includegraphics[width=\linewidth]{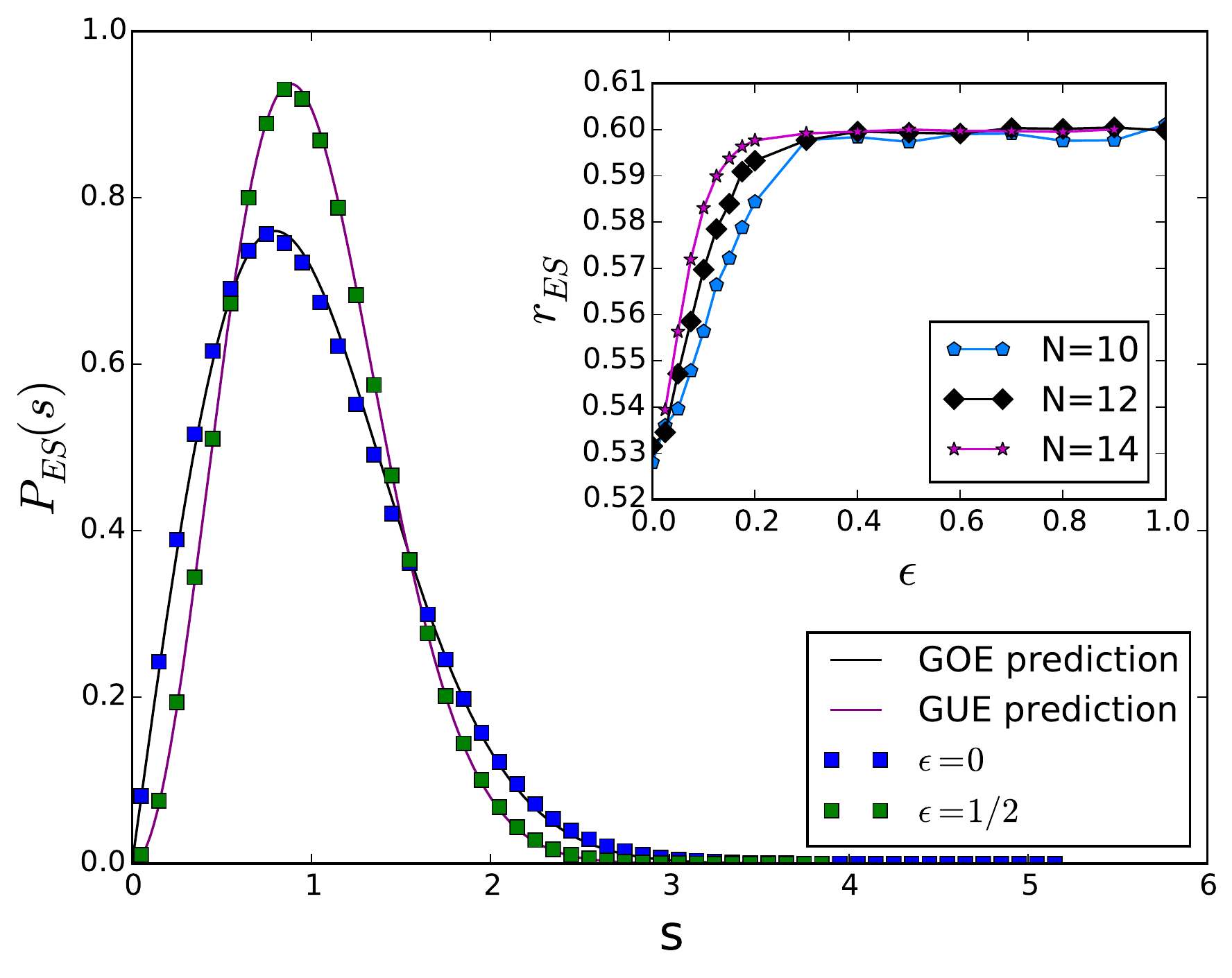}
\caption{ Demonstration of how the entanglement level statistics depends on the origin of time. The blue dots are the entanglement spacings for a time-reversal symmetric origin of time, they satisfy a GOE distribution, which is the same as the Floquet energy spacings. The green dots are for the TR-breaking drive of Eq. \ref{TwoBangEvolution}, they follow a GUE distribution.  All data was taken for $N=14$, $h=6$. (Inset) The level spacing ratio for various system sizes, which takes the GOE value only when $\epsilon=0$ (the TR symmetric point).}\label{changing_ensemble}
\end{figure}

We note that while the Floquet eigenenergies are independent of the `origin of time' used when defining the Floquet unitary operator, the eigenstates of that unitary operator (and hence the entanglement spectrum) {\it do} depend on the choice of origin of time. The Floquet unitary operator with a general choice of origin of time can be written as:
\begin{equation}
U(\tau,\epsilon) = \exp(-i H_1 \frac{\tau+\epsilon}{4}) \exp(-i H_2 \frac{\tau}{2}) \exp(-i H_1 \frac{\tau-\epsilon}{4}).\label{TwoBangEvolutionShiftedOrigin}
\end{equation}
This evolution operator is time-reversal symmetric at $\epsilon=0$ and identical to Eq.~\ref{TwoBangEvolution} for $\epsilon=\tau$. The eigenstates $|\varphi_i (\tau,\epsilon)\rangle$ are related for different values of $\epsilon$ by the unitary transformation
\begin{equation}
|\varphi_i (\tau,\epsilon)\rangle= \exp(-i H_1 (\epsilon'-\epsilon)/4) |\varphi_i (\tau,\epsilon')\rangle
\end{equation}
We clearly see that the entanglement spectrum of $|\varphi_i (\tau,\epsilon)\rangle$ depends on the choice of $\epsilon$ unless $H_1$ does not couple the two regions $A$ and $B$ defining the bipartition. In our setup, Eq.~\ref{H1Integrable} has one term that prevents such a separation. Thus the choice of origin of time will have a dramatic consequence on our entanglement spectrum analysis. We show the level spacing distributions of the entanglement spectrum for the ordinary TR-breaking protocols in Fig.~\ref{changing_ensemble}. We can see that at the TR symmetric point $\epsilon=0$, the entanglement level spacings are GOE, just like the Floquet energy distributions. (The distribution is GOE and not COE because the entanglement Hamiltonian is Hermitian) The inset of Fig. ~\ref{changing_ensemble} shows how the level spacing ratio $r_{\rm ES}$ depends on $\epsilon$, we see that $r_{\rm ES}$ is only GOE for a TR-symmetric choice of origin of time. 

\begin{figure}[htb]
\includegraphics[width=\columnwidth]{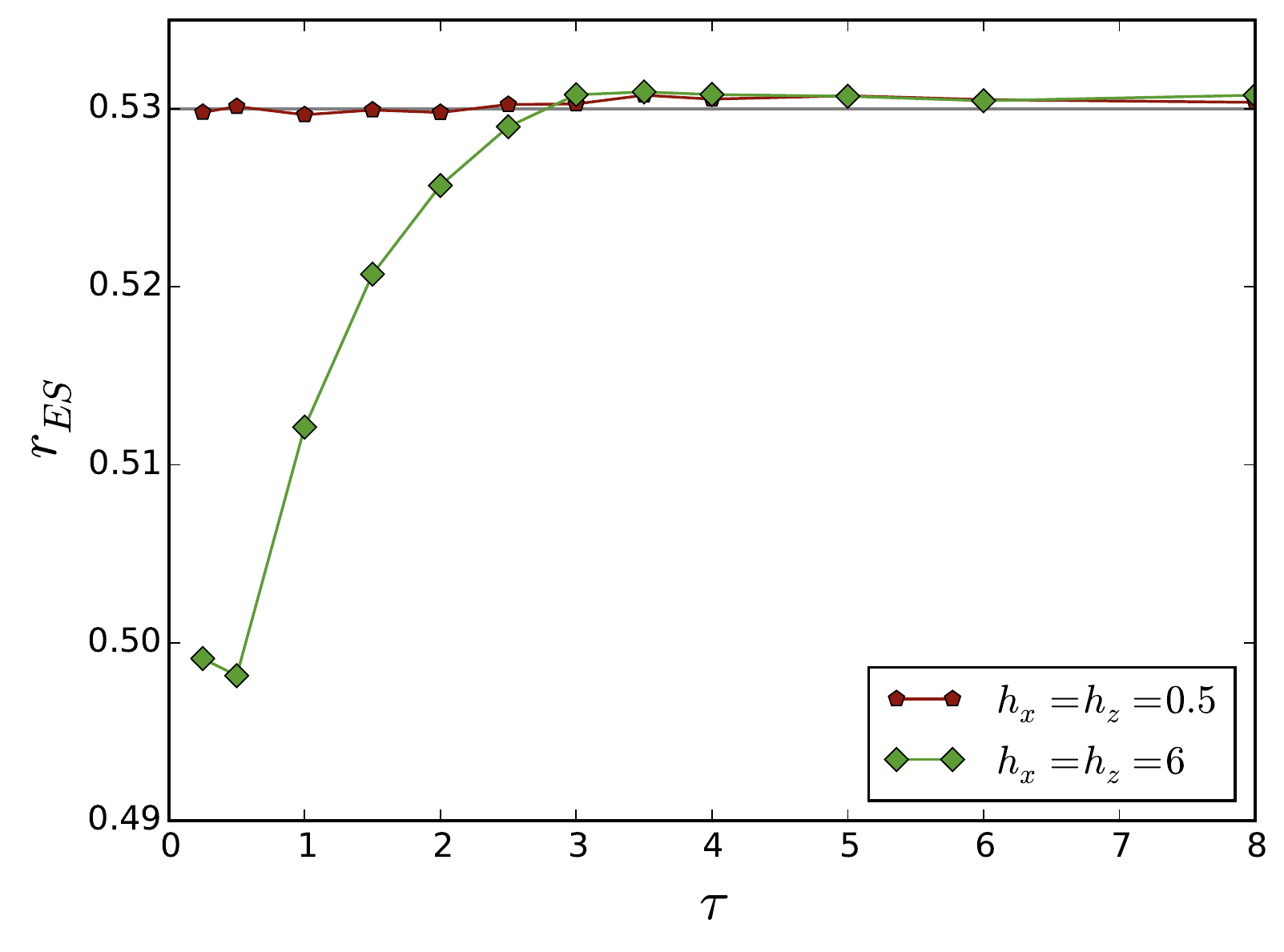}
\caption{ Level spacing ratios $r_{\rm ES}$ for the entanglement spectrum for the same model as that shown in Fig.~\ref{EnergyDOSLevelStatIntegrableCOE}a, but with a time-reversal symmetric origin of time. The grey line shows the predicted value for the GOE.}\label{floquet_ent_r}
\end{figure}

Fig.~\ref{floquet_ent_r} shows the level spacing ratio $r_{\rm ES}$ for the entanglement spectrum with a time-reversal symmetric drive, for the system whose Floquet energy spacing ratios are shown in Fig.~\ref{EnergyDOSLevelStatIntegrableCOE}a. Similar to the Floquet energy case, the entanglement level spacing ratios are GOE at large $\tau$, and at small $\tau$ are GOE for small disorder and Semi-Poisson for large disorder. Note however that at unlike in the Floquet energy case there no `dip' in the spacing ratios between the two $\tau$ regimes.

\section{Conclusion}

We have studied the entanglement spectrum in the thermal and localized phases of Hamiltonian and Floquet systems, revealing a wealth of information about the entanglement structure. In the thermal phase of Hamiltonian systems, the entanglement spectrum is governed by random matrix theory, in accordance with a particularly strong form of the eigenstate thermalization hypothesis. The conclusion appears independent of dimensionality. In Floquet systems, the entanglement spectrum is controlled by corrections to ETH, but is still governed by random matrix theory. However, the random matrix ensemble governing the entanglement spectrum in Floquet systems can be different to that governing the spectrum of Floquet eigenphases. In particular, the entanglement spectrum will follow GUE level statistics unless the energy spectrum is COE, the drive has an inversion center {\it and} we place the origin of time at the inversion center (note that in Floquet systems the entanglement spectrum depends on the choice of origin of time). 

Meanwhile in the localized phase, we observe that the entanglement spectrum has level statistics that follow a semi-Poisson distribution, unlike the energy spectrum, which follows Poisson level statistics. This semi-Poisson form is observed in one and two dimensional Hamiltonian systems, and in Floquet localized phases, and thus appears to be a universal property of the entanglement structure of the localized phase. However, it is {\it not} seen in non-interacting localized phases i.e. the semi-Poisson statistics appear to be a consequence of the interacting nature of the localized phase. We have conjectured an explanation for the semi-Poisson statistics of the entanglement spectrum in terms of many body resonances. 

Semi-Poisson statistics are generally considered a diagnostic of criticality, and the entanglement structure even deep in the localized phase thus appears to show a `residual criticality' that has not hitherto been appreciated. We have identified a `two peak' structure in the entanglement spectrum of the localized phase, and have pointed out that information about the level repulsion and semi-Poisson statistics is contained mainly in the `high entanglement energy' part of the entanglement spectrum, with the low entanglement energy states (that control entanglement entropy and Renyi entropies) showing little sign of level repulsion. Thus it appears that this residual criticality in the entanglement structure is likely to be invisible to entropy measures, and could only have been revealed by an analysis of the full entanglement spectrum using diagnostics inspired by random matrix theory. Whether the high energy entanglement spectrum deep in the MBL phase retains a memory of any other properties of the critical point (such as other critical exponents), and if so how this information could be extracted, is an intriguing question that we leave to future work. We note that semi-Poisson statistics are also a diagnostic of pseudo-integrability \cite{Berry, Bogomolny}, and our results thus suggest that the entanglement Hamiltonian in the many body localized phase should be {\it pseudointegrable} (i.e. not integrable but also not chaotic), an observation that may open new lines of attack on the localized phase. 

Finally, we have verified that performing microcanonical averages of level statistics for entanglement energies gives essentially the same results as calculating level statistics for the entanglement spectrum of a single eigenstate. While this is expected in the thermal phase (a corollary of ETH), it is a surprise in the localized phase, where eigenstates are known to vary dramatically in their properties (temperature chaos). Nonetheless it appears that the statistical properties of the entanglement structure do not vary much from one eigenstate to the next, such that microcanonical averaging may be employed as a useful tool when studying level statistics of entanglement spectra even in the many body localized phase. 

Thus, a study of the entanglement spectrum has revealed a rich and hitherto unsuspected structure to the pattern of entanglement in both thermal and localized phases. Further exploration of such ideas may open a new line of attack on the localized and delocalized phases, and the intervening phase transition. We leave further exploration of these ideas to future work. 

{\it Acknowledgments:} We acknowledge fruitful discussions with C. Von Keyserlingk, T. Neupert, A. Chandran, S. Kourtis, R. Bhatt, B.A. Bernevig, S. Capponi, N. Laflorencie and, especially, David Huse. N.R. was supported by the Princeton Global Scholarship. S.G. was supported by DOE-BES Grant DE-SC0002140.

\appendix

\section{Different ways to average}
\begin{figure}[htb]
\includegraphics[width=\linewidth]{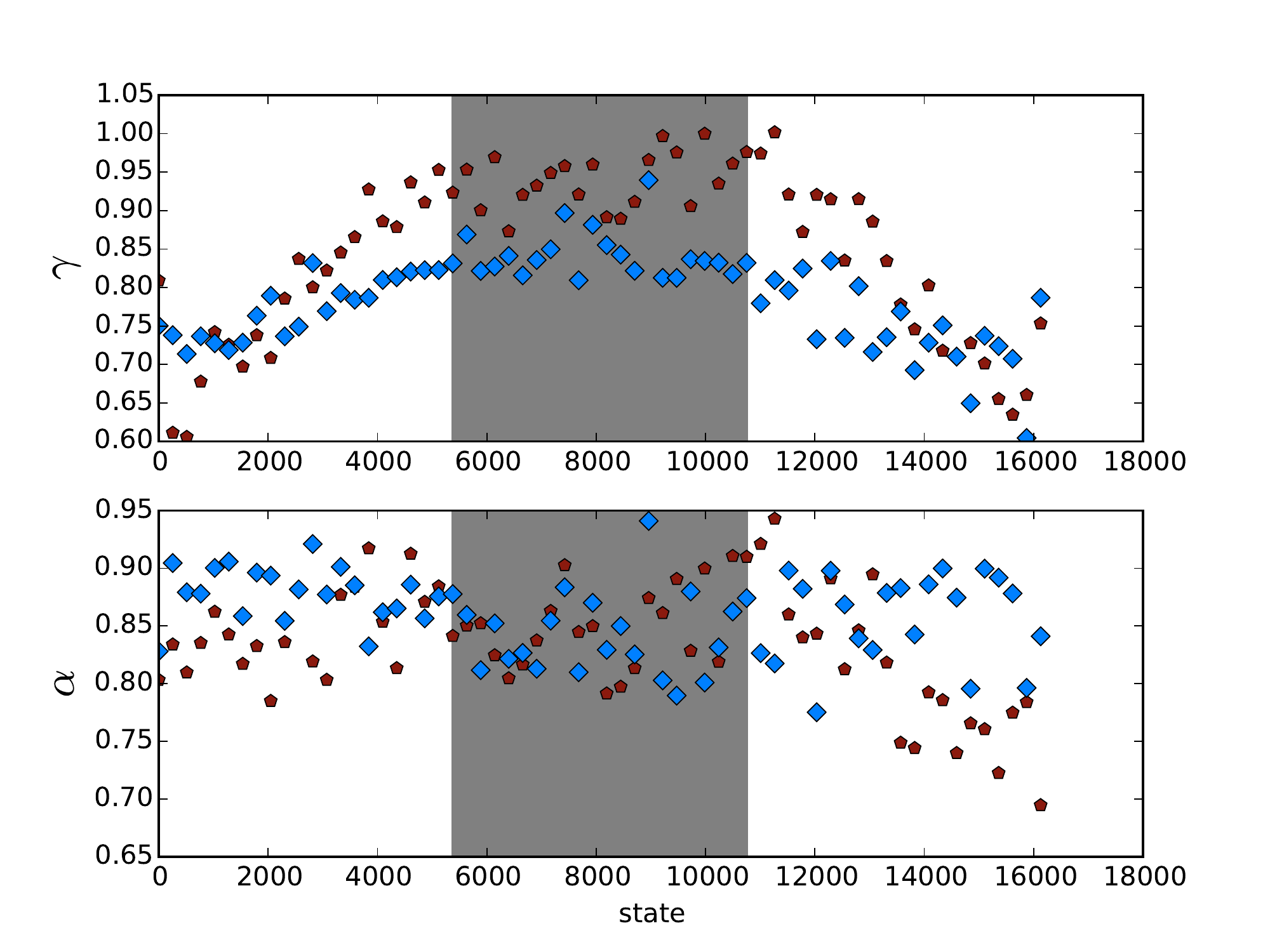}
\caption{This plot was obtained by extracting the values of $\alpha$ and $\gamma$ for $256$ energy eigenstates at a time. The horizontal axis shows the index of the first state on the set of $256$. We see that the at values at small and large energy are different from that in the middle of the spectrum. The region which we average over in the main text is shown in gray, we can see the exponents are relatively constant in this region. The different colored points correspond to different realizations of disorder, we see that changing the disorder realization can affect the extracted exponents, at least for the finite size systems available to our numerics. Data was taken with $N=14$, $h_x=h_z=12$, $h_y=0$.\label{whole_spectrum}}
\end{figure}

In the main text at small sizes we had the problem that there were not enough entanglement states to draw concrete conclusions about their level statistics. We found that averaging over a number of entanglement states gave the same results as a single state, and therefore this is the approach that we used throughout the paper. Specifically, we averaged over all states lying within the middle third of the energy spectrum, but one might wonder how this energy window was chosen. We support this choice in Fig.~\ref{whole_spectrum}. In this figure we estimate the exponents $\alpha$ and $\gamma$ from Eq.~\ref{DistributionGammaBeta}, as a function of location in the energy spectrum, which we computed by averaging over $256$ states. We can see that at the high and low ends of the spectrum, slightly different values for $\gamma$ are obtained. The shaded region shows the middle third of states which were averaged over in the main text, we can see that in this region the exponents are relatively constant.

\begin{figure}[htb]
\includegraphics[width=\linewidth]{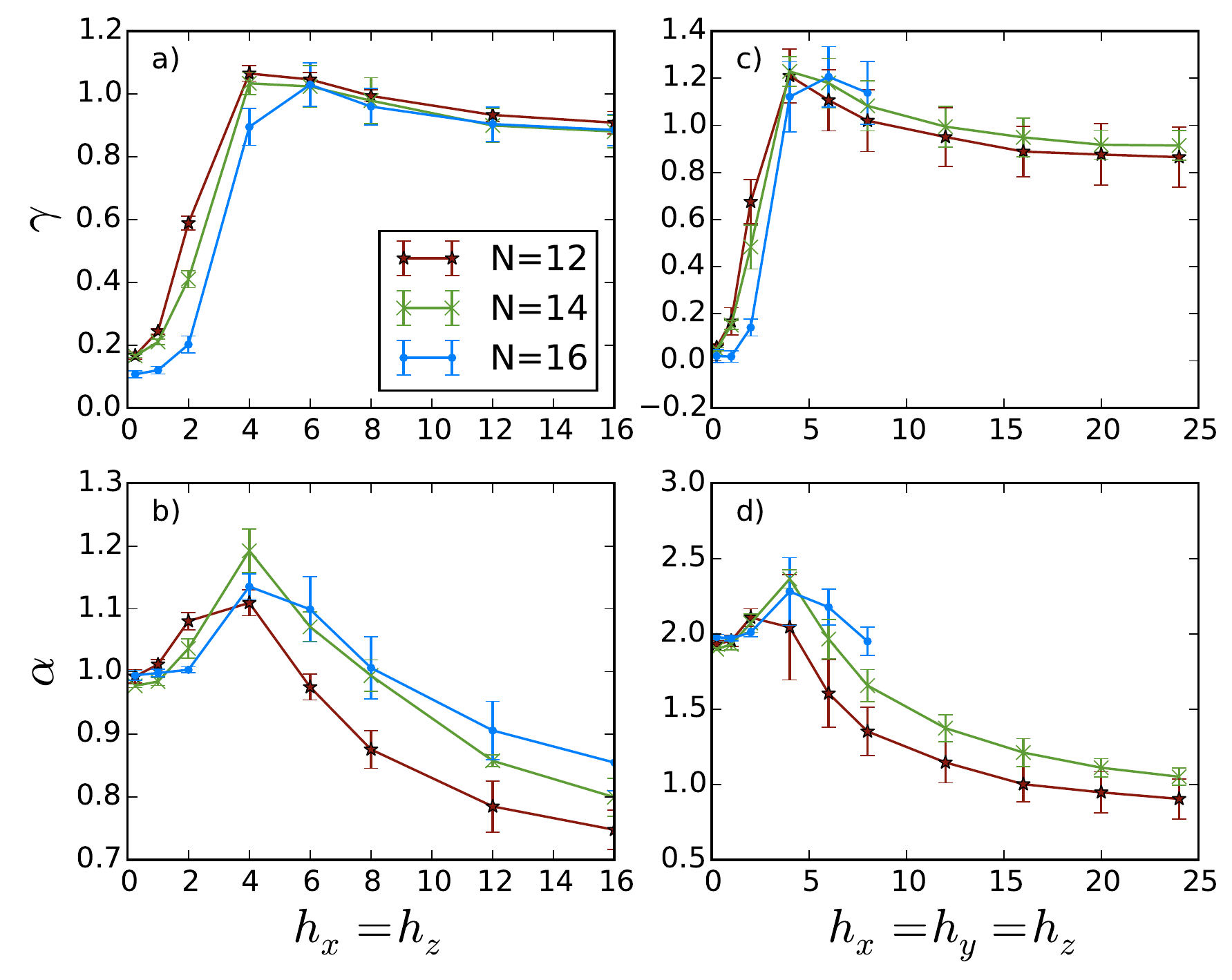}
\caption{ Fitted values of (a) $\gamma$ and (b) $\alpha$ (as defined in Eq.~\ref{DistributionGammaBeta}) for the ES, for multiple sizes and averaged over multiple realizations of disorder (16 ROD for $N=12$, $4$ for $N=14,16$). Data was taken for the GOE case with $h_x=h_z=h$, $h_y=0$. For $N=12,14$ the entanglement spectra in each disorder realization was averaged over the one-third of energy levels in the middle of the spectrum, for $N=16$ the spectra were averaged over $1000$ energy levels near the middle of the energy spectrum. $\alpha$ and $\gamma$ were extracted for each realization of disorder, and error bars are the variance of the different values.\label{GOE_exp} (c,d) Same as (a,b) but in the unitary symmetry class $h_x=h_y=h_z$. In this case the entanglement eigenvalues are smaller so that at $h>6$ we run into problems with machine precision at $N=16$. \label{GUE_exp}}
\end{figure}

In the main text, all quantities are computed for only a single realization of disorder (ROD), but we can ask how averaging over disorder could improve our data. In Fig.~\ref{whole_spectrum} the different symbols correspond to different ROD, and we can see that different realizations can give slightly different results. Note that this is a finite-size effect, at larger sizes the system would better self-average and differences between ROD would disappear. In Fig~\ref{GUE_exp} we average over ROD to improve our estimates of the exponents $\alpha$ and $\gamma$. The improved estimates allow us to see how these quantities vary with system size. The results support our hypothesis that in the limit of large $N$ and large $h$, $\alpha=\gamma=1$. 

\bibliography{eslevelstat.bib}

\end{document}